\documentclass[11pt, a4paper]{article}

\usepackage[utf8]{inputenc}
\usepackage[T1]{fontenc}
\usepackage{lmodern}

\usepackage[margin=1in]{geometry}
\usepackage{setspace}
\usepackage{titlesec}
\titleformat*{\section}{\large\bfseries\boldmath}
\titleformat*{\subsection}{\normalsize\bfseries\boldmath}
\titleformat*{\paragraph}{\normalsize\bfseries\boldmath}
\usepackage{amsmath, amssymb, amsthm, amsfonts, mathtools, xspace, dsfont}
\usepackage{booktabs} %,microtype,enumitem,needspace}
\allowdisplaybreaks

\usepackage{graphicx}
\usepackage{tikz}
\usetikzlibrary{arrows.meta,decorations.pathreplacing}
\usepackage{xcolor}
\usepackage[most]{tcolorbox}
\usepackage[colorinlistoftodos, textsize=footnotesize]{todonotes}
\newcounter{DannyComment}

\usepackage[authoryear,round]{natbib}

\definecolor{darkblue}{rgb}{0.0, 0.0, 0.55}
\usepackage[pagebackref=true, colorlinks=true, allcolors=darkblue]{hyperref}
\renewcommand*{\backref}[1]{}
\renewcommand*{\backrefalt}[4]{%
    \ifcase #1%
    \or #2%
    \else #2%
    \fi%
}

\makeatletter
\newtheoremstyle{smallcaps}
  {}% Space above
  {}% Space below
  {\itshape}% Body font (Italic)
  {}% Indent amount
  {\scshape}% Theorem head font (Small Caps)
  {.}% Punctuation after theorem head
  { }% Space after theorem head
  {}% Theorem head spec

\makeatother
\theoremstyle{smallcaps}
\newtheorem{theorem}{Theorem}[section]
\newtheorem{lemma}[theorem]{Lemma}

\newtheorem{claim}[theorem]{Claim}

\newcommand{\pr}[1]{\operatorname{Pr}\left[#1\right]}
\newcommand{\prpar}[1]{\operatorname{Pr}[#1]}
\newcommand{\prsub}[2]{\operatorname{Pr}_{#1}\left[#2\right]}
\newcommand{\prsubpar}[2]{\operatorname{Pr}_{#1}[#2]}

\newcommand{\ex}[1]{\mathbb{E}\left[#1\right]}

\newcommand{\exsubpar}[2]{\mathbb{E}_{#1}[#2]}

\newcommand{\eps}{\epsilon}

\newcommand{\bbZ}{\mathbb{Z}}

\newcommand{\opt}{\mathrm{OPT}}
\newcommand{\ind}{\mathds{1}}
\newcommand{\UU}{\mathcal{U}}
\newcommand{\PP}{\mathcal{P}}
\newcommand{\psim}[1]{\psi_{#1}(\theta)}
\newcommand{\pihat}{\hat\pi}

\title{\boldmath\textbf{Assortment Optimization under the Mallows Model: \\ 
Independent-Utility Choice
Equivalence \\
and Its Algorithmic Implications}}
\author{Danny Segev\thanks{School of Mathematical Sciences, Tel Aviv University,
Tel Aviv 69978, Israel. Email: \texttt{segevdanny@tauex.tau.ac.il}. Research supported by Israel Science Foundation grant 767/26.}}
\date{}

\begin{document}
\maketitle
\begin{abstract}
\noindent The Mallows model is a well-established construct in preference modeling, describing
random rankings through their Kendall--Tau distance from a reference permutation.
Its compact parameterization and interpretable treatment of preference
uncertainty have made it influential across statistics, machine learning, probability,
and revenue management. In the latter domain, recent literature has studied the embedding of Mallows-driven choice into assortment optimization. Early approximation schemes in this context require stylized structural simplifications or restrictive parametric assumptions (D{\'e}sir, Goyal, Jagabathula, and Segev; {\em Operations Research}~2021). In the absence of these levers, a highly involved quasi-polynomial-time
approximation scheme has subsequently been proposed (Rieger and Segev; {\em Mathematical Programming}~2024), leaving wide open the question of whether near-optimal assortments
can be computed in true polynomial time, with our analytical understanding and algorithmic toolbox severely lacking.

\medskip\noindent The main contribution of this focused technical paper resides in closing these gaps by uncovering a striking independent-utility structure
within Mallows choices, and turning this representation into a polynomial-time approximation scheme. Our central structural characterization explicitly constructs a single collection of mutually
independent utilities, reproducing Mallows choice probabilities
simultaneously for all offered sets. This finding is particularly counterintuitive, since we concurrently demonstrate that the full Mallows
ranking distribution cannot, in general, be generated via 
independent utilities. This exact choice equivalence connects pending Mallows-related challenges to the recent approximation framework of Bowers et al.~(FOCS~2025) for assortment optimization under independent random utilities. We make this connection algorithmically feasible through a
finite-support reduction that produces distributions of polynomial
size while controlling revenue loss uniformly over all assortments.
Fused together, these ingredients lead to an $O( n^{ \tilde{O}( 1 / \eps^2 ) } )$-time approximation scheme for assortment
optimization under the  Mallows model, while allowing arbitrary product prices, concentration
parameters, and modal rankings.
\end{abstract}

\medskip

\thispagestyle{empty}

% --- Table of Contents & List of Todos ---
\newpage
\thispagestyle{empty}
\tableofcontents
% \newpage
% \thispagestyle{empty}
% \listoftodos[List of Comments]
\newpage

% sections %%%%%%%%%%%%%%%%%%%%%%%%%%%
\setcounter{page}{1}
\section{Introduction}\label{sec:intro}

Since its introduction in Mallows' seminal work \citeyearpar{mallows1957}, the Mallows model has long served as a  versatile framework for describing random rankings around a
common reference order. Its defining principle is that rankings become exponentially
less likely as their dissimilarity to this order increases, with a single
concentration parameter controlling the dispersion around this preference. While we cannot do justice to the surrounding literature, it is still worth noting that the Mallows model and its generalizations have extensively been  studied as tools
for the statistical analysis of rank data
\citep{fligner1986,marden1995}. Their applications in machine learning
and information retrieval include rank aggregation
and consensus estimation \citep{lebanon2002,meila2007}, learning from
pairwise comparisons \citep{lu2014}, and recommendation based on
incomplete preferences \citep{vitelli2018}.

Moving beyond a single reference order, mixtures of Mallows distributions
allow different segments within a given population to exhibit distinct preference
patterns. Recovering these latent segments and their preferences has
motivated a substantial line of research \citep{awasthi2014, liu2018}. Alongside questions of statistical
inference, the Mallows model has attracted interest as a family of nonuniform
random permutations, and its limiting behavior, subsequence statistics, and
cycle structure have been investigated in depth
\citep{starr2009,bhatnagar2015,gladkich2018}. Connections to Markov-chain
sampling \citep{diaconis2000} and likelihood-based interpretations of
voting \citep{conitzer2005} further illustrate the diversity of questions
that arise from this captivating model.

\paragraph{Discrete choice and assortment optimization.} In light of its compact parameterization and interpretability, the Mallows model offers a natural framework for assortment
optimization, connecting customers' preferences over products
to the purchase probabilities and expected revenues associated
with different offerings. In such settings, given representative customers who select their most preferred available alternative, we wish to decide which products should be offered to maximize expected revenue. On this front,  \citet{farias2013} developed a nonparametric
approach that utilizes distributions over preference lists to predict
assortment revenues from limited choice data. Related approaches to
representing and learning preferences include the repeated insertion
model of \citet{doignon2004} and the treatment of partially observed
rankings by \citet{lebanon2008}. From a decision-making standpoint,
\citet{desir2016,desir2021} initiated the algorithmic study of assortment
optimization under Mallows-based choice. This line of work encompasses distinct algorithmic
approaches, with nuanced differences in their assumptions,
guarantees, and limitations that will be thoroughly discussed in
Section~\ref{sec:prior}.

Concurrently with these advances, recent work has further expanded the conceptual
modeling and estimation methodology required to connect Mallows-based choice to
observed customer behavior. Among others, these contributions include the partial-order estimation
framework of \citet{jagabathula2018}, together with its  Mallows likelihood
interpretation, and the estimation
and prediction methods for cross-sectional and panel data of
\citet{jagabathula2025}. 
Additional studies addressed settings in which
preferences are expressed through shorter rankings. For example, \citet{chierichetti2018}
introduced Mallows models for \mbox{top-$k$} lists, and \citet{haddadan2025}
devised algorithmic approaches for sampling, computing choice probabilities, and parameter
learning under a generalized top-$k$ model. Yet another extension of the Mallows model captures ranked choices
via the reverse major index; here, \citet{feng2022,feng2025} mainly investigated the
choice law induced by this alternative distance measure and its preference estimation. 

\paragraph{Outline.} To smoothly transition from this introductory background to our particular topic, Section~\ref{sec:model} formalizes the Mallows choice model and presents its induced assortment
optimization problem.
Section~\ref{sec:prior} then reviews the computational landscape around this subject, culminating in three  multifaceted open questions motivating the current paper. Finally, Section~\ref{sec:contributions} presents our main contributions and briefly touches upon their design principles and technical innovation.

\subsection{The Mallows model and assortment optimization} \label{sec:model}

\paragraph{Rankings and the Mallows distribution.} To shed rigorous light on the choice mechanism we consider, let $\UU=[n]$ be our underlying universe of products, and let
$0$ designate the ever-present no-purchase option. Together, these two components form the overall set of alternatives
$\UU^+=\UU\cup\{0\}$. A ranking of the latter set is simply a bijection
$\sigma:\UU^+\to[n+1]$, where $\sigma(i)$ stands for the position of
alternative $i$, with smaller positions indicating stronger preferences. Letting $\PP_{n+1}$ be the collection of all such rankings, the Mallows distribution is specified by a modal ranking
$\omega\in\PP_{n+1}$ and a concentration parameter $\theta\ge0$. Given these ingredients, the probability assigned to each ranking $\sigma$ is exponentially decreasing in its
Kendall--Tau distance $d(\sigma,\omega)$ from $\omega$, counting the alternative pairs that are ordered differently by the two rankings, i.e.,
\[
 d(\sigma,\omega) ~~=~~
 \sum_{\substack{i,j\in\UU^+:\\i<j}}
 \ind\big[(\sigma(i)-\sigma(j))(\omega(i)-\omega(j))<0\big].
\]
Specifically, each ranking $\sigma\in\PP_{n+1}$ occurs with probability
\begin{equation}\label{eq:mallows}
 \prsub{(\omega,\theta)}{\sigma}
 ~~=~~
 \frac{e^{-\theta \cdot d(\sigma,\omega)}}{\psi(\theta)},
\end{equation}
where the normalizing constant above is
$\psi(\theta)=\prod_{i=1}^{n+1}
\frac{1-e^{-i\theta}}{1-e^{-\theta}}$. By inspecting this representation, one immediately notices that the concentration parameter $\theta$ controls how strongly the Mallows distribution
favors rankings close to $\omega$. For example, at $\theta=0$, all rankings are
equally likely, with $\psi(0)= \lim_{\theta \to 0} \psi(\theta) = (n+1)!$. At the other extreme, as $\theta$ tends to infinity,
the resulting distribution concentrates on the modal ranking $\omega$.

\paragraph{Choice probabilities.} We proceed by explaining how the Mallows model governs choice in assortment optimization contexts. Here, given an assortment $S\subseteq\UU$ of offered products, our representative customer can always decide against making a purchase, meaning that
her available alternatives are captured by $S^+=S\cup\{0\}$. The choice
process we consider first draws a random ranking $\sigma$ according to equation~\eqref{eq:mallows}
and then selects the highest-ranked alternative in $S^+$. To conveniently express this rule, noting that alternative $i$ is selected
when $\sigma(i)=\min_{j\in S^+}\sigma(j)$, its choice probability $\pi(i,S)$ can be written
by averaging over the Mallows distribution, namely,
\[
  \pi(i,S) ~~=~~ \sum_{\sigma\in\PP_{n+1}}
  \prsub{(\omega,\theta)}{\sigma} \cdot
  \ind\left[\sigma(i)=\min_{j\in S^+}\sigma(j)\right].
\]

\paragraph{The assortment optimization problem.} In this setting, each product $i\in\UU$ is associated with a fixed nonnegative price $r_i$, and we
set $r_0=0$ to reflect the absence of revenue when no purchase is made.
Then, the choice mechanism described above determines the expected
revenue of any assortment $S$ as $R(S) = \sum_{i\in S}r_i\cdot\pi(i,S)$. Our objective is to compute an assortment whose expected revenue is maximized. As explained later on, throughout this paper, we consider the unconstrained single-segment
formulation, in which all customers follow the same Mallows distribution,
and every subset of $\UU$ is a feasible assortment. For computational purposes, our  input description consists of the modal ranking $\omega$, encoded as a
permutation, together with the concentration parameter $\theta$
and the product prices $r_1,\ldots,r_n$, each specified as a
rational number.

\paragraph{Additional notation.} To avoid cumbersome expressions, it will be convenient to denote the overall number of alternatives by $N=n+1$, as well as to make use of the dispersion parameter $q=e^{-\theta}$. For every integer $m\ge0$, we introduce the
$m$-normalizer
$\psim{m}=\prod_{i=1}^{m}[i]_q$, where
$[i]_q=\sum_{\ell=0}^{i-1}q^\ell  = \frac{ 1 - q^i }{ 1 - q }$, with the convention that $\psim{0}=1$. In particular, equation~\eqref{eq:mallows} has $\psi(\theta) = \psim{N}$, and we suppress the argument
$\theta$ whenever it is fixed by writing $\psi_m = \psim{m}$. In addition, for
every nonempty set $A\subseteq\UU^+$ and alternative $i \in A$, we define $\pihat(i,A)=\prsubpar{(\omega,\theta)}{
 \sigma(i)=\min_{j\in A}\sigma(j)}$. This notation allows us to discuss choices from arbitrary sets of
alternatives, with $\pi(i,S)=\pihat(i,S^+)$, thereby recovering the assortment optimization
model. 

\subsection{Known results and open questions}\label{sec:prior}

We next review the computational landscape of assortment optimization
under Mallows-based choice. Due to their vastly different characteristics, our discussion distinguishes between a single
Mallows segment and mixtures thereof, while tracking previously employed
assumptions, required to efficiently compute near-optimal assortments.
Of particular importance are the number of segments involved, their concentration
parameters, and the largest-to-smallest price ratio. 

\paragraph{Inapproximability results.} At present, provable hardness results are available only in regard to mixtures of arbitrarily many Mallows segments.
In this extension, each segment $k\in[K]$ has its own modal ranking
$\omega_k$ and concentration parameter $\theta_k$.
A customer first selects one of these segments, with respective probabilities $\{ \lambda_k \}_{k \in [K]}$, and then draws
a random ranking from the corresponding Mallows distribution. Somewhat unsurprisingly, the strong inapproximability of this setting emerges from the observation that, by sufficiently increasing the concentration parameters $\{ \theta_k \}_{k \in [K]}$,
each Mallows segment can approximate a deterministic ranking to arbitrary accuracy, as pointed out by \citet[Sec.~1.2]{rieger2024}. For distributions over fixed rankings,
\citet[Thm.~1]{aouad2018} showed that it is NP-hard to approximate the optimal expected revenue within a factor of $O(K^{1-\eps})$, for any fixed
$\epsilon\in(0,1)$, and the above-mentioned relation translates these inapproximability bounds to mixtures of Mallows segments. Still, this relationship does not establish intractability results for any fixed number of segments, which is a major open question on which no progress has been made to date.

\paragraph{First approximation schemes.}
The algorithmic study of Mallows-based assortment optimization was initiated
by \citet{desir2016}, who developed polynomial-time procedures for
evaluating choice probabilities in this context along with a compact mixed-integer formulation
for computing revenue-maximizing assortments. Building on this foundation, \citet{desir2021} designed three
approximation schemes, all applicable to mixtures of Mallows segments. Nevertheless, the last sentence is missing the phrase ``polynomial time'' for a very good reason, as each of these approaches depends on stylized structural simplifications or restrictive parametric assumptions:
\begin{itemize}
    \item
    When the no-purchase option appears last in every modal ranking,
    there exists a $(1-\eps)$-approximate assortment consisting of 
    only $O( \frac{ 1 }{ \eps })$ products.
    Enumerating assortments of this cardinality therefore yields an $O(Kn^{O(1/\eps)})$-time approximation scheme.

    \item
    To allow arbitrary modal rankings,
    an interval-enumeration approach exploits the concentration of
    rankings around their respective modes. This approach leads to a
    $(1-\eps)$-approximation in $O((
        \frac{n}{\eps\hat\theta}
        \cdot\frac{r_{\max}}{r_{\min}}
        )^{O(K/\hat\theta)})$ time, where $r_{\max}$ and $r_{\min}$ designate the largest and smallest
    positive prices, and $\hat\theta=\min_{k\in[K]} \min \{ \theta_k, \frac{ 1 }{ 2 } \}$. 

    \item
    A further enhancement eliminates the above-mentioned dependence on concentration by
    searching over structured assortments defined through price
    and choice probability thresholds. Here, the resulting running time is $O( n^{ (\frac{ 1 }{ \eps} \log( \frac{ nr_{\max} }{ r_{\min} }) )^{ O(K) } } )$, which is double-exponential in the number of segments $K$.
\end{itemize}
It is worth pointing out that the assumption underlying the first scheme is essential to its
small-assortment argument.
Indeed, when the no-purchase option occupies an arbitrary modal
position, even a constant-factor approximation may require an
assortment of linear size \citep[Thm.~5]{desir2021}.
While the remaining schemes accommodate such instances, their running
times retain a strong dependence on concentration or on the range of possible prices.
Thus, even for a single segment, these algorithmic developments leave a significant gap between
near-optimality under specific assumptions and polynomial-time
approximation in full generality.

\paragraph{The unrestricted QPTAS.}
For a single Mallows segment, \citet{rieger2024} overcame these
parametric restrictions through the design of a quasi-polynomial-time approximation scheme (QPTAS). Specifically, for any $\eps > 0$, they devised a deterministic $O(n^{O_\eps(\log n)})$-time procedure for computing an assortment whose expected revenue is within a factor of  $1-\eps$ of optimal, allowing arbitrary product prices, concentration
parameters, and modal rankings. At a high level, their approach combines a block decomposition of the modal ranking with a separate partition into $O( \frac{\log n }{ \eps } )$ intervals defining $P$-constrained assortments, and coordinates their selection through dynamic programming. One  apparent downside of these advances is their involved technical nature, both algorithmically and analytically. Moreover, due to searching over structured descriptions of $P$-constrained assortments, the logarithmic running time exponent specified above seems unavoidable when following this pathway.

\paragraph{Main open questions.}
This state of affairs gives rise to a broad spectrum of interrelated questions. The motivating inquiries listed below are mostly concerned with whether a polynomial-time approximation scheme (PTAS) is within reach, the right algorithmic toolbox to tackle Mallows-based assortment optimization,
and novel representations of the underlying choice mechanics:
\begin{itemize}
    \item \emph{PTAS without structural or parametric restrictions?} For a wide range of combinatorial optimization problems, quasi-polynomial-time approximation schemes have been state-of-the-art for quite a while; see, e.g., minimum-weight triangulation \citep{remy2009}, maximum-weight independent set of axis-parallel rectangles \citep{adamaszek2013}, precedence-constrained unit-size makespan scheduling on $O(1)$ identical machines \citep{garg2018}, and tree search with nonuniform query costs \citep{dereniowski2017}.
    Does the current setting fall into this family, or can we exploit its unique features to develop an assumption-less approximation scheme in true polynomial time?     

    \item \emph{New algorithmic toolbox for Mallows-based choice?}
    Can recent developments in algorithm design, perhaps appearing unrelated to Mallows-based choice, provide an efficient route to address assortment optimization in this context? As previously explained, earlier approaches either rely on specific structural and parametric assumptions, or suffer from built-in quasi-polynomial running time, suggesting that we should be looking for significantly different ideas.    

    \item \emph{Optimization-oriented representation?}
    While Mallows choice probabilities can be efficiently evaluated,
    their definition through random rankings makes the interactions
    among offered products difficult to analyze.
    Can a single collection of independent utilities reproduce these
    probabilities, simultaneously for all offered sets?
    Moreover, for algorithmic purposes, can such a representation be captured by 
    distributions with polynomially many support points while
    preserving near-optimal expected revenues?
\end{itemize}

\subsection{Main results and technical highlights}\label{sec:contributions}

The main contribution of this focused technical paper resides in resolving these questions in the affirmative, by uncovering a well-hidden independent-utility structure within Mallows choices
and turning this representation into a polynomial-time approximation scheme. In particular, our central structural characterization explicitly constructs a single collection of mutually
independent utilities that jointly reproduce the Mallows choice probabilities,
simultaneously for all offered sets.
This finding is particularly striking, since we concurrently demonstrate that the Mallows
ranking distribution cannot, in general, be generated via independent utilities.
By preserving the expected revenue of every assortment, our
representation connects Mallows-based choice to the recent approximation framework of
\citet{bowers2025} for independent random utilities.
We make this connection algorithmically feasible through a
finite-support reduction that produces distributions of polynomial
size while controlling revenue loss uniformly over all assortments.
As formally stated below, fused together, these ingredients lead to a polynomial-time approximation scheme for assortment
optimization under the  Mallows model, while allowing arbitrary product prices, concentration
parameters, and modal rankings.

\begin{theorem}\label{thm:main}
Assortment optimization under the Mallows model admits a polynomial-time approximation scheme. For any accuracy level  $\eps\in(0,1)$, our algorithm can be implemented in $O( n^{ \tilde{O}( 1 / \eps^2 ) } )$ time.
\end{theorem}

% It is worth mentioning that, to arrive at the most digestible presentation, we do not make any attempts to optimize the running time exponent beyond $\tilde{O}( 1 / \eps^2 )$. In addition, we measure running time in a unit-cost real-arithmetic model,
% with exact arithmetic, comparisons, and evaluations of
% $\exp$, $\log$, and the floor function. Thus, our running-time
% bounds count operations independently of the binary precision
% needed to represent their arguments and output.

\paragraph{An unexpected route through independent utilities.} As alluded to above, our work creates a conceptual bridge between the Mallows model and the  polynomial-time approximation scheme of
\citet{bowers2025} for assortment optimization under independent
random utilities. The specific version that will be useful in this context permits the
no-purchase utility to  be random, provided that it remains independent of all product utilities. This setting is explicitly covered in Section~\mbox{IV-C3} of their paper, in conjunction with its treatment of the outside option
in Section~\mbox{IV-C2}.
Combining their Corollary~36 with the running-time analysis
in Section~\mbox{III-F} yields the following guarantee.

\begin{theorem}
\label{thm:external}
Consider $n$ products with fixed nonnegative prices and mutually
independent utilities for all products and the no-purchase option.
Suppose that each utility distribution is explicitly specified by at
most $m$ mass points and their corresponding probabilities. Then, for any $\eps\in(0,1)$, we can compute a deterministic 
assortment whose expected revenue is within a factor of $1-\eps$ of optimal,
using $O((\frac{nm}{\eps})^{O(1/\eps^2)})$ arithmetic operations.
\end{theorem}

At first sight, however, connecting this theorem to the Mallows model seems impossible, given its independence assumption.
The most direct approach would be to generate a Mallows ranking
by sorting mutually independent utilities. Unfortunately, as we demonstrate in Section~\ref{sec:nonrum}, such a representation
simply does not exist with $N\ge4$ alternatives and dispersion $0<q<1$, even with a fixed priority
rule for breaking ties.
This obstruction is structural: Independent utilities make
events defined for  disjoint pairs of alternatives independent,
whereas the Mallows ranking distribution does not satisfy this
property. That said, for assortment optimization, reproducing the entire
ranking distribution is unnecessary, since its expected revenue function depends only on the probabilities of selecting
each alternative from a given offered set.
Our pivotal insight is that a single collection of independent
utilities can reproduce these probabilities simultaneously for
every offered set, despite the impossibility of reproducing the
full Mallows ranking distribution.
This distinction forms the missing link that brings
Theorem~\ref{thm:external} into play.

\paragraph{Technical overview.} To this end, our overall approach synthesizes two main ingredients: An exact representation
of Mallows choices through independent utilities, followed by an efficiently implementable
reduction to finitely supported distributions. Section~\ref{sec:overview} provides a technical overview of how these ingredients combine
with Theorem~\ref{thm:external} to yield our polynomial-time approximation scheme, tracking the expected revenue losses along the way and deriving the overall running time. We proceed by outlining these ingredients in greater detail.

\paragraph{Milestone I: Representing Mallows choices by independent utilities.} Our first ingredient is an exact representation of Mallows-based
choice probabilities through mutually independent utilities $\{ U_i \}_{ i \in \UU^+ }$, created by applying alternative-specific
deterministic shifts to independent copies of a common noise
distribution. We prove that selecting the largest realized utility reproduces the Mallows choice probabilities
for every offered set, meaning that this construction preserves the expected revenue of every assortment. Section~\ref{sec:representation} establishes this
equivalence by expressing both choice probabilities as sums
indexed by subsets of unavailable alternatives.
On the Mallows side, each subset specifies which
alternatives precede the selected alternative in a sampled
ranking.
On the independent-utility side, the same subsets arise by
expanding the expression for the winning probability.
Evaluating the resulting terms shows that the two expansions
coincide, revealing an independent representation at the level
of choices even though none exists at the level of rankings.

\paragraph{Milestone II: Creating finitely supported utilities.} Our second ingredient addresses an inherent computational hurdle of this exact representation, which associates each utility $U_i$ with an unbounded support, whereas
Theorem~\ref{thm:external} requires explicitly specified
distributions with polynomially many mass points. We overcome this obstacle through two successive operations: Grouping and truncation. First, we group nearby score values into intervals whose width
depends on the concentration parameter $\theta$ and the desired
accuracy $\eps$. The particular form of our noise distribution allows us to
multiplicatively bound the resulting change in each choice
probability. Subsequently, we truncate the tails of these grouped distributions, while 
controlling the additive revenue loss through an explicit
lower bound on the optimal expected revenue. Taken as a whole, these operations produce independent
utility distributions with polynomially sized supports.
The accompanying revenue bounds we obtain hold uniformly over all
offered sets, allowing near-optimal assortments with respect to the resulting
finite instance to be transferred back to the original Mallows
instance.
Section~\ref{sec:finite} establishes these bounds, completing
the connection between our exact choice representation and
the approximation scheme of \citet{bowers2025}.
\section{Mallows Rankings Cannot Be Generated by Independent Utilities}\label{sec:nonrum}

Before constructing independent utilities to express Mallows-based choice probabilities in terms of random utility maximization, we examine
whether such a representation could recover the full
ranking distribution. The latter is a stronger requirement, since for assortment
optimization, only the probabilities $\{ \pi(i,S) \}_{i \in S}$ of selecting each available product affect the expected revenue of any assortment $S$. The purpose of this section is to show why these two
requirements lead to different answers.

\subsection{Choice equivalence and the ranking obstruction}

\paragraph{Choice under independent utilities.} An independent random utility model consists of mutually independent
real-valued random variables $\{ U_i \}_{i\in\UU^+}$. Their decreasing order, combined
with any tie-breaking rule based on a fixed deterministic total order, induces a random
ranking. We say that two ranking laws are choice equivalent when, for every
nonempty set $A\subseteq\UU^+$ and  alternative $i\in A$, they assign the same probability to $i$ being best in $A$. While equality of ranking laws clearly implies choice equivalence, the converse can fail, since the choice probabilities
for individual offered sets need not uniquely determine the
underlying ranking law.
This distinction is closely related to the identification
problem studied by \citet{turansick2022}, who provides necessary and sufficient conditions for uniqueness of the distribution over rankings that generates a given system of choice probabilities.

\paragraph{Impossibility result.} As formally stated below, we prove that Mallows rankings cannot be represented by independent utilities in general. Our proof begins with a small-scale example, comparing the probabilities of two events defined by disjoint pairs of alternatives. Here, independent utilities would force these events to be independent as well; however, a direct
calculation under the Mallows law leads to a strictly positive covariance. With this outcome in place, we generalize our analysis to any number of underlying alternatives via the repeated insertion method.

\begin{theorem}\label{thm:nonrum}
When $N\ge4$ and $0<q<1$, the Mallows ranking law
cannot be generated by independent utilities, even with a fixed priority
rule for breaking ties.
\end{theorem}

\subsection{Proof of Theorem~\ref{thm:nonrum}} \label{subsec:proof_thm_nonrum}

\paragraph{Construction for $N=4$.} Let us first consider a universe consisting of four alternatives, with the modal ranking $\omega = abcd$ written from most to least
preferred. We now examine the events
${\cal E}_{ac}=\{\sigma(a)<\sigma(c)\}$ and
${\cal E}_{bd}=\{\sigma(b)<\sigma(d)\}$, where $a$ precedes $c$
and $b$ precedes $d$ in the sampled ranking $\sigma$, respectively. To evaluate their probabilities under the Mallows model, let
$W({\cal E})=\sum_{\sigma\in {\cal E}}q^{d(\sigma,\omega)}$ denote the unnormalized
weight of an event ${\cal E}$, meaning that $W({\cal E})= \psi_4 \cdot \prpar{{\cal E}}$ by representation~\eqref{eq:mallows} of our ranking
probabilities. Then, the weights of ${\cal E}_{ac}$, ${\cal E}_{bd}$, and ${\cal E}_{ac} \cap {\cal E}_{bd}$ are given by
\[ W({\cal E}_{ac}) ~~=~~ W({\cal E}_{bd}) ~~=~~ 1+3q+3q^2+3q^3+2q^4 \qquad \text{and} \qquad W({\cal E}_{ac}\cap {\cal E}_{bd}) ~~=~~ 1+3q+q^2+q^3. \]
To better understand where these expressions are coming from, let us consider the rankings in ${\cal E}_{ac}\cap {\cal E}_{bd}$, which are exactly $abcd$, $abdc$, $acbd$, $bacd$, $badc$, and $bdac$. Their inversion counts
are $0,1,1,1,2,3$, leading to a total weight of $1+3q+q^2+q^3$. Here, we are indeed seeing $q^0$ with a coefficient of $1$, corresponding to a single ranking with zero inversions, $q^1$ with a coefficient of $3$, corresponding to three rankings with one inversion, and so on. For either of the
events ${\cal E}_{ac}$ and ${\cal E}_{bd}$, the numbers of rankings with $0,1,2,3,4$ inversions are
$1,3,3,3,2$, respectively, and there are no rankings with more inversions, leading to $1+3q+3q^2+3q^3+2q^4$. Normalizing these weights by $\psi_4 =(1+q)(1+q+q^2)(1+q+q^2+q^3)$, we obtain
\begin{align}
\pr{{\cal E}_{ac}\cap {\cal E}_{bd}}-\pr{{\cal E}_{ac}} \cdot \pr{{\cal E}_{bd}} ~~&=~~ \frac{ W({\cal E}_{ac}\cap {\cal E}_{bd}) }{ \psi_4 } - \frac{ W({\cal E}_{ac}) }{ \psi_4 } \cdot \frac{W({\cal E}_{bd}) }{ \psi_4 } \nonumber \\
& =~~\frac{q^3(1-q)^2}{(1+q+q^2) \cdot \psi_4} \nonumber \\
& >~~0, \label{eqn:not_independent}
\end{align}
noting that this strict inequality holds for every $0<q<1$.

By contrast, under an independent utility model with a fixed
priority rule, whether $a$ precedes $c$ is determined solely
by $(U_a,U_c)$, including when these utilities are equal.
Likewise, whether $b$ precedes $d$ is determined solely by
$(U_b,U_d)$. Since these pairs involve
disjoint sets of independent variables, we must have $\prpar{{\cal E}_{ac}\cap {\cal E}_{bd}} = \prpar{{\cal E}_{ac}} \cdot \prpar{{\cal E}_{bd}}$, contradicting inequality~\eqref{eqn:not_independent}.

\paragraph{The repeated insertion method.} To address the broader case of $N > 4$ alternatives, it is instructive to briefly discuss the well-known repeated insertion method for sampling permutations
from the Mallows distribution. Readers are referred to \citet{doignon2004} and \citet{lu2014} for an in-depth discussion.
In this procedure, we insert the alternatives in ${\cal U}^+$ into
a single ranking, one after another, according to their order in
the modal ranking $\omega$. When considering the $i$-th ranked
alternative, it is inserted into the ranking constructed thus far
at a random position $j\in\{1,\ldots,i\}$, where position $1$ is the most
preferred and position $i$ is the least preferred. Independently
of previous insertions, position $j$ is selected with
probability $ \frac{q^{i-j}}{[i]_q}$, recalling that $[i]_q=\sum_{\ell=0}^{i-1}q^\ell$. Conversely, each final ranking
uniquely determines its sequence of insertion positions.
Moreover, inserting the $i$-th alternative at position $j$ creates
exactly $i-j$ inversions relative to $\omega$, while preserving all
comparisons among previously inserted alternatives. Consequently,
by multiplying these terms, we end up with a  probability of
$\frac{ q^{d(\sigma,\omega)} }{ \prod_{i=1}^N[i]_q } = \frac{e^{-\theta \cdot d(\sigma,\omega)}}{\psi(\theta)}$ for every final ranking
$\sigma$, matching equation~\eqref{eq:mallows}.

\paragraph{Extended construction for $N>4$.}
To lift our construction to arbitrarily large $N$, let $B=\{a,b,c,d\}$ be the set of alternatives occupying positions $1,2,3,4$ in the modal
ranking $\omega$, respectively. Consider generating a Mallows
ranking via the repeated insertion procedure. Then, its first four steps
coincide with applying this procedure to $B$ alone, with the modal ranking $abcd$ and dispersion parameter $q$. Consequently,
at the end of the fourth step, these alternatives are ranked according to the four-alternative Mallows distribution.
Every subsequent step inserts another alternative without changing
the relative order of those already present. Hence, the relative
ranking induced on $B$ by the completed procedure has this same
distribution. If the full ranking law were generated by mutually
independent utilities $\{U_i\}_{i\in\UU^+}$, its restriction to $B$
would be generated by the mutually independent utilities
$U_a,U_b,U_c,U_d$, with the same fixed priority rule restricted
to $B$. Such a construction would provide an independent utility representation of
the four-alternative Mallows law, contradicting the $N=4$ case discussed above.
\section{Approximation Scheme: Technical Overview}\label{sec:overview}

In what follows, we describe the nuts and bolts of our approximation scheme by reducing Mallows-based assortment
optimization to an independent random utility maximization instance while incurring negligible revenue loss en route. Along these lines, Section~\ref{subsec:milestone_1} constructs independent utilities that
reproduce the Mallows choice probabilities for every offered set,
thereby preserving the expected revenue of every assortment.
Section~\ref{subsec:reduction_finite} then reduces their unbounded supports to polynomially
many points through grouping and truncation, while controlling
the resulting optimality loss uniformly over all
assortments. Section~\ref{subsec:alg-step-3} applies the approximation scheme of
\citet{bowers2025} to this assortment optimization instance and transfers
its performance guarantees back to the original problem. For ease of exposition, the complete proofs of our exact representation and its finite-support
bounds are provided in Sections~\ref{sec:representation}
and~\ref{sec:finite}.

\subsection{Milestone I: Mallows choices via independent utilities} \label{subsec:milestone_1}

We begin by generating independent integer-valued random scores along with a rank-based tie-breaking rule, jointly reproducing
choice probabilities under the Mallows model for every offered set. This representation will initially be defined over an unbounded support, thereby feeding our finite-support
reduction, whose finer details are provided in milestone~II.

\paragraph{Boundary cases.} For the remainder of this discussion to be well-defined, we will assume that $q \in (0,1)$. Luckily, the case $q=1$ as well as the limiting case $q=0$ are straightforward to address from an assortment optimization perspective. First, by circling back to Section~\ref{sec:model}, we know that when $q=0$, the Mallows distribution places its  entire probability mass on the modal ranking $\omega$. As such, a product can be sold
only when it precedes the no-purchase option in $\omega$, meaning that offering a highest-price such product alone is optimal; if none exists, the optimal expected revenue is clearly zero. At the other extreme, all $N!$ rankings are equally likely at $q=1$, implying that with respect to any assortment $S \subseteq \UU$, each alternative in $S^+$ has probability $\frac{ 1 }{ |S|+1 }$ of being chosen. Therefore, our expected revenue function takes the simple form
$R(S) = \sum_{i\in S} \frac{ r_i }{ |S|+1 }$. Here, for any fixed cardinality $k \leq n$, this expression is maximized by the $k$ most expensive products, meaning that there are only $O(n)$ assortments to be considered.

\paragraph{Defining random utilities/scores.} For simplicity of analysis, the upcoming construction will interleave ``utilities'' and ``scores'', with the first notion generating pairwise disjoint supports for all alternatives, and with the second allowing overlaps, thus requiring a predefined tie-breaking rule. Our first step assigns each alternative an independent random score
whose deterministic part reflects its modal position. To this end, for
every $i\in\UU^+$, let $Z_i$ be an integer-valued random
variable, such that for all $k\in\bbZ$,  
\begin{equation}\label{eq:noise-cdf}
 F_{ Z_i }(k) ~~=~~\pr{Z_i\le k}~~=~~\frac{1}{1+q^k}.
\end{equation}
Since $q \in (0,1)$, it is easy to verify that $F_{Z_i}$ is a valid cumulative distribution function. Here, the noise variables $\{ Z_i \}_{ i\in\UU^+ }$ are independent and identically distributed, without any relation to modal positions. Next, we obtain the scores $\{ X_i \}_{ i\in\UU^+ }$ by subtracting the respective modal positions, that is, each alternative $i\in\UU^+$ is associated with a score of $X_i=Z_i-\omega(i)$, noting that 
\begin{equation} \label{eqn:CDF_Xi}
F_{X_i}(k) ~~=~~ F_{ Z_i }(k+\omega(i)) ~~=~~\frac{1}{1+q^{k+\omega(i)}}.    
\end{equation}
Once these scores are generated, from any offered set, we select the alternative with the largest score,
breaking ties in favor of the smaller modal rank.

As mentioned above, it will occasionally
be convenient to express this choice mechanism through the rational-valued utilities
\begin{equation}\label{eq:real-utility}
 U_i~~=~~X_i-\frac{\omega(i)}{N+1}.
\end{equation}
Since the scores $\{ X_i \}_{i\in\UU^+}$ are integer-valued, the  offsets $\{ \frac{\omega(i)}{N+1} \}_{i\in\UU^+}$
preserve every strict score comparison and resolve all ties according to
the prescribed rule. In particular, distinct alternatives cannot have
equal utilities.

\paragraph{Choice equivalence.} The main structural property of this construction, formally stated below, is that it recovers
the Mallows choice probabilities for every set of available alternatives. The proof of this result is presented in Section~\ref{sec:representation}, where we compute the winning probability under the independent utilities $\{ U_i \}_{i\in\UU^+}$ and
match it through an appropriate decomposition of the corresponding Mallows event.

\begin{theorem} \label{thm:representation}
For every nonempty set $A\subseteq\UU^+$ and  alternative $i\in A$,
\[
 \pr{U_i=\max_{j\in A}U_j}~~=~~\pihat(i,A).
\]
\end{theorem}

\paragraph{Implications.} Consequently, once
the random utilities $\{ U_i \}_{i\in\UU^+}$ are specified, every assortment's expected revenue can be evaluated in terms of these
independent utilities. Moving forward, we make use of $R_X(S)$ to denote the expected revenue of an assortment
$S\subseteq\UU$ when selecting the alternative with
the largest score among $S^+$, breaking ties in favor of the
smaller modal rank. Similarly, $R_U(S)$ will designate the analogous expected
revenue, obtained by selecting the unique utility maximizer
among $S^+$. Since these two rules select the same alternative at every score realization, for every assortment $S\subseteq\UU$, we have
\[
 R_X(S)
 ~~=~~ R_U(S)
 ~~=~~ \sum_{i\in S}r_i \cdot \pr{U_i=\max_{j\in S^+}U_j} ~~=~~ \sum_{i\in S}r_i \cdot \pihat(i,S^+)
 ~~=~~ R(S),
\]
where the next-to-last equality follows from Theorem~\ref{thm:representation}. Thus, we can study our original assortment optimization problem
through these independent scores/utilities.

\subsection{Milestone II: Creating finitely supported utilities}
\label{subsec:reduction_finite}

Our next step consists of replacing the unbounded scores $\{X_i\}_{i\in\UU^+}$ by
independent scores with polynomially many support points, which is a fundamental requirement to employ the approximation scheme of
\citet{bowers2025} along the lines of Theorem~\ref{thm:external}. At present, it is still unknown whether such supports suffice to represent Mallows choice probabilities in an exact way, and we therefore aim at designing ``approximate'' replacement scores, eventually leading to negligible loss in optimality. This objective will be achieved by first grouping nearby scores,
which results in a controlled multiplicative distortion of the expected revenue, and then truncating their tails, with an additive loss that will be appropriately bounded as well.

\paragraph{Lower-bounding $\opt$.} To measure the truncation-related additive loss we will incur, let $\opt = \max_{S\subseteq\UU}R(S)$ designate the optimal expected revenue with respect to the original
assortment optimization problem. The next claim, whose proof is provided in Appendix~\ref{app:proof_clm_LB_B_OPT}, shows that this optimum can be lower-bounded by $\frac{ q^{N-1} }{ N } \cdot r_{\max}$. The key observation behind our proof is that a maximum-price product alone yields revenue $r_{\max}$
whenever it appears first in a Mallows-sampled ranking, with the repeated insertion method showing that this event has probability
at least $\frac{ q^{N-1} }{ N }$, regardless of the product's position in the
modal ranking.

\begin{claim} \label{clm:LB_B_OPT}
$\opt \geq \frac{ q^{N-1} }{ N } \cdot r_{\max}$.   
\end{claim}

\paragraph{Alteration 1: Grouping nearby scores.} Given an accuracy parameter $\eps\in(0,1)$, we coarsen the integer-valued scores $\{ X_i \}_{i\in\UU^+}$ by partitioning the integer line into blocks of common width $\Delta=1+\lfloor\frac{\eps}{\theta}\rfloor$ and setting the grouped scores $Y_i=\lfloor\frac{X_i}{\Delta}\rfloor$. Figure~\ref{fig:finite-support}(a) illustrates this grouping
operation with $\Delta=4$. Clearly, $\{ Y_i \}_{i\in\UU^+}$ are supported by the entirety of $\bbZ$, and we should therefore operate according to a predefined tie-breaking rule: Once again, when selecting the alternative with
the largest score, ties are broken in favor of the
smaller modal rank.

To examine the consequences of coarsening $\{ X_i \}_{i\in\UU^+}$ into $\{ Y_i \}_{i\in\UU^+}$, let $R_Y$ denote the expected revenue function induced
by these grouped scores. The following result, whose proof is presented in Section~\ref{subsec:proof_lem_coarse}, shows that up to a multiplicative factor of $e^{\pm \eps}$, this function matches our original expected revenue function $R_X$ on every assortment.  

\begin{figure}[htpb!]
\centering
\resizebox{0.98\linewidth}{!}{%
\begin{tikzpicture}[
    font=\small, >=Stealth,
    score/.style={draw=blue!65!black, fill=blue!5,
                  minimum width=8mm, minimum height=7mm},
    endpoint/.style={draw=teal!80!black, fill=teal!6,
                     minimum width=10mm, minimum height=7mm}
]
% (a) Grouping, with h=4.
\node[anchor=west] at (0,3.3)
    {(a) $Y_i=\lfloor X_i/\Delta\rfloor$};
\node[anchor=west] at (0,2.15) {$X_i$};
\node[anchor=west] at (0,0.76) {$Y_i$};
\node at (1.05,2.15) {$\cdots$};
\node at (15.1,2.15) {$\cdots$};

\foreach \g in {0,1,2} {
    \draw[draw=black!20, fill=blue!4]
        ({1.86+4.2*\g},1.8)
        rectangle ({5.89+4.2*\g},2.5);
    \foreach \j in {0,1,2,3} {
        \pgfmathtruncatemacro{\value}{4*\g+\j-4}
        \node at ({2.3+1.05*(4*\g+\j)},2.15) {$\value$};
    }
    \pgfmathtruncatemacro{\groupvalue}{\g-1}
    \draw[->, thick, blue!65!black]
        ({3.875+4.2*\g},1.65) --
        ({3.875+4.2*\g},1.17);
    \node[score] at ({3.875+4.2*\g},0.76)
        {$\groupvalue$};
}
\draw[black!20] (0,0.1) -- (15.4,0.1);

% (b) Truncation.
\node[anchor=west] at (0,-0.45)
    {(b) $\bar{Y}_i=
      \max\{-\Gamma,\min\{Y_i,\Gamma\}\}$};
\node[anchor=west] at (0,-1.7) {$Y_i$};
\node[anchor=west] at (0,-3.1) {$\bar{Y}_i$};
\draw[black!25] (1.35,-1.7) -- (14.85,-1.7);

\foreach \x/\lab in {
    2.1/{-\Gamma-2},4/{-\Gamma-1},5.9/{-\Gamma},
    8.1/{0},10.3/{\Gamma},12.2/{\Gamma+1},14.1/{\Gamma+2}} {
    \node at (\x,-1.15) {$\lab$};
    \fill[black!50] (\x,-1.7) circle (1.2pt);
}
\foreach \x in {7,9.2} {
    \node at (\x,-1.55) {$\cdots$};
    \node at (\x,-3.1) {$\cdots$};
}
\foreach \x/\lab in {5.9/{-\Gamma},8.1/{0},10.3/{\Gamma}} {
    \node[endpoint] at (\x,-3.1) {$\lab$};
    \draw[->, black!50] (\x,-1.8) -- (\x,-2.72);
}
\foreach \x/\target in {2.1/5.9,4/5.9,12.2/10.3,14.1/10.3} {
    \fill[orange!85!black] (\x,-1.7) circle (1.2pt);
    \draw[->, thick, orange!85!black]
        (\x,-1.8) -- (\target,-2.72);
}
\end{tikzpicture}%
}
\caption{Grouping integer scores into blocks of width $\Delta=4$, followed by
truncation to $[-\Gamma,\Gamma]$, where each tail is mapped to the nearest endpoint.}
\label{fig:finite-support}
\end{figure}

\begin{lemma}\label{lem:coarse}
For every assortment $S\subseteq\UU$, 
\[  e^{-\eps} \cdot R_X(S)~~\le~~ R_Y(S)~~\le~~ e^\eps \cdot  R_X(S). \]
\end{lemma}

While its underlying arguments are somewhat involved, this bound follows from the observation that once all competing scores are fixed, alternative $i$ wins precisely when
its own score crosses a certain threshold. Measured on the
original score scale, grouping shifts this threshold by at most
$\Delta-1$.
The corresponding conditional winning probabilities will therefore
differ by a factor between $q^{\Delta-1}$ and $q^{-(\Delta-1)}$.
Since $\Delta=1+\lfloor\frac{\eps}{\theta}\rfloor$, we know that $q^{\Delta-1}=e^{-\theta(\Delta-1)}\ge e^{-\eps}$ and $q^{-(\Delta-1)}=e^{\theta(\Delta-1)}\le e^\eps$, implying that averaging over the competing scores allows us to derive these multiplicative
bounds for each alternative's choice probability, which translate to analogous bounds for the expected revenue of any assortment.

\paragraph{Alteration 2:  Truncating grouped scores.}
We next retain only a bounded interval of grouped scores, assigning all
remaining probability mass to its two endpoints. Specifically, letting
$\Gamma=\lceil \frac{ 2N^2 }{ \eps^2 }\rceil$, each score $Y_i$ will keep its value when falling within $[-\Gamma,\Gamma]$; otherwise, it will be mapped to the nearest endpoint of this interval. In other words, $\bar Y_i=\max\{-\Gamma,\min\{Y_i,\Gamma\}\}$. As illustrated in Figure~\ref{fig:finite-support}(b), scores
within this interval remain unchanged, while each tail is
mapped to the corresponding endpoint. As such, each truncated score belongs to
$\{-\Gamma,-\Gamma+1,\ldots,\Gamma\}$, thus taking only $O( \Gamma )$ possible values.

To account for the potential revenue loss due to considering $\{ \bar{Y}_i \}_{i\in\UU^+}$ in place of $\{ Y_i \}_{i\in\UU^+}$, let
$R_{\bar{Y}}$ be the expected revenue function after truncation. The next claim, whose proof is provided in Section~\ref{subsec:proof_lem_clip}, shows that the difference between the expected revenues before and after
truncation is negligible in comparison to $\opt$, for any possible assortment. 

\begin{lemma}\label{lem:clip}
For every assortment $S\subseteq\UU$,
\[ | R_{\bar{Y}}(S)-R_Y(S)|~~\le~~2\eps \cdot \opt. \]
\end{lemma}

\paragraph{The truncated  utilities.} In summary of the preceding discussion, it remains to describe the resulting distributions of $\{ \bar{Y}_i \}_{i \in \UU^+ }$. Here, for every alternative $i \in \UU^+$, the probability mass function of $\bar{Y}_i$ can be directly computed from the cumulative distribution function of $Y_i$. Since $Y_i=\lfloor\frac{X_i}{\Delta}\rfloor$, the latter is given by $F_{Y_i}(y)  = F_{ X_i }( \Delta(y+1)-1 )$ for every $y \in \bbZ$. With this connection, the distribution of the truncated score $\bar{Y}_i$ has mass $F_{Y_i}( -\Gamma )$ at $-\Gamma$, mass $F_{Y_i}(y) - F_{Y_i}(y-1)$ at every $-\Gamma<y<\Gamma$, and mass $1-F_{Y_i}( \Gamma-1)$ at
$\Gamma$. 

\subsection{Final algorithm and approximation guarantee} \label{subsec:alg-step-3}

\paragraph{Black-box application.} We conclude this section by specifying the assortment optimization instance to which the approximation scheme of \citet{bowers2025} is applied. To avoid incompatibility between their specific way of resolving ties and our modal-order tie-breaking rule, we prevent ties altogether by associating  every alternative $i\in\UU^+$ with random utility $\bar U_i
 =
 \bar Y_i-\frac{\omega(i)}{N+1}$. As in our original representation, these offsets
preserve strict comparisons between scores and resolve equal
scores in favor of the smaller modal rank. For this reason, the
utilities $\{\bar U_i\}_{i\in\UU^+}$ have a unique maximizer
over every nonempty set of alternatives, and their induced
expected revenue function is precisely $R_{\bar Y}$.

All in all, the instance created  consists of the 
product prices $r_1, \ldots, r_n$, along with the explicit  distributions of $\{\bar U_i\}_{i\in\UU^+}$. In particular, the marginal for each alternative $i$ has
support points
$y- \frac{ \omega(i) }{ N+1 }$, for $y\in\{-\Gamma,\ldots,\Gamma\}$, whose respective probabilities have already been specified
in Section~\ref{subsec:reduction_finite}.
Moreover, the utilities $\{\bar U_i\}_{i\in\UU^+}$ remain mutually independent, since
grouping, truncation, and deterministic offsets are separately applied to each original score. The no-purchase option is always available, has revenue zero, and
retains its own independent random utility. These properties place our instance within the exact setting
of Theorem~\ref{thm:external}, including its allowance for a
random no-purchase utility. When this approximation scheme is employed with accuracy
$\eps$, we obtain an assortment $\bar S\subseteq\UU$ with an expected revenue of 
\begin{equation} \label{eqn:eps_approx_bar_S}
   R_{\bar Y}(\bar S)
 ~~\ge~~
 (1-\eps) \cdot \max_{S\subseteq\UU}R_{\bar Y}(S). 
\end{equation}

In terms of computational efficiency, an explicit construction of all $N$ finite marginals requires $O(N\Gamma)$ operations. Subsequently, invoking the approximation scheme of Theorem~\ref{thm:external} takes $O( ( \frac{ N\Gamma }{\eps})^{O( 1 / \eps^{2})} )$ operations. By recalling that $N=n+1$ and $\Gamma=\lceil \frac{ 2N^2 }{ \eps^2 }\rceil$, we end up with a total running time of $O( ( \frac{ N\Gamma }{\eps})^{O( 1 / \eps^{2})} ) = O( n^{ \tilde{O}( 1 / \eps^2 ) } )$, precisely as stated in Theorem~\ref{thm:main}.

\paragraph{Overall approximation guarantee.} We proceed by transferring the above-mentioned bounds with respect to the finitely supported scores $\{ \bar{Y}_i \}_{i \in \UU^+ }$ to the Mallows expected revenue function. For this purpose, let $S^*$ be an optimal assortment for our original Mallows
instance, with an expected revenue of $R(S^*) = \opt$. We argue that the assortment $\bar{S}$ must be near-optimal, by observing that it garners an expected revenue of
\begin{align}
R( \bar{S} ) ~~&=~~ R_X( \bar{S} ) \label{eqn:final_apx_1} \\
~~&\geq~~ e^{-\eps} \cdot R_Y( \bar{S} ) \label{eqn:final_apx_2} \\
~~&\geq~~ e^{-\eps} \cdot \left( R_{\bar{Y}}( \bar{S} ) - 2\eps \cdot \opt \right) \label{eqn:final_apx_3} \\
~~&\geq~~ e^{-\eps} \cdot \left( (1-\eps) \cdot R_{\bar{Y}}( S^* ) - 2\eps \cdot \opt \right) \label{eqn:final_apx_4} \\
~~&\geq~~ e^{-\eps} \cdot \left( (1-\eps) \cdot R_{{Y}}( S^* ) - 4\eps \cdot \opt \right) \label{eqn:final_apx_5} \\
~~&\geq~~ e^{-\eps} \cdot \left( (1-\eps) \cdot e^{-\eps} \cdot R_{X}( S^* ) - 4\eps \cdot \opt \right) \label{eqn:final_apx_6} \\
~~&\geq~~ (1 -7\eps) \cdot \opt . \label{eqn:final_apx_7}
\end{align}
Here, equality~\eqref{eqn:final_apx_1} holds since $R( S ) = R_X( S )$ for every assortment $S \subseteq \UU$, as shown in Section~\ref{subsec:milestone_1}. Inequalities~\eqref{eqn:final_apx_2} and~\eqref{eqn:final_apx_3} follow from Lemmas~\ref{lem:coarse} and~\ref{lem:clip}, respectively. Inequality~\eqref{eqn:final_apx_4} is an immediate implication of equation~\eqref{eqn:eps_approx_bar_S}, which shows in particular that $R_{\bar Y}(\bar S) \geq (1-\eps) \cdot R_{\bar Y}(S^*)$. Inequalities~\eqref{eqn:final_apx_5} and~\eqref{eqn:final_apx_6} again follow from Lemmas~\ref{lem:clip} and~\ref{lem:coarse}, respectively. Finally, inequality~\eqref{eqn:final_apx_7} holds since
$R_X(S^*) = R(S^*) = \opt$, $(1-\eps) \cdot e^{-2\eps} \geq 1-3\eps$,
and $e^{-\eps}\leq 1$ for $\eps\in(0,1)$.

\section{Exact Choice Equivalence}\label{sec:representation}

The current section is dedicated to proving Theorem~\ref{thm:representation}, which states that for every offered set, the independent utilities $\{ U_i \}_{i \in \UU^+}$
and the Mallows ranking select each alternative with precisely the same
probability. Formally, we argue that $\prpar{U_i=\max_{j\in A}U_j} = \pihat(i,A)$ for every  nonempty set $A\subseteq\UU^+$ and  alternative $i\in A$. Toward this objective, in Section~\ref{subsec:summation-identity}, we derive a summation
identity that connects our logistic score distribution to the
Mallows normalizers. In Sections~\ref{subsec:mallows-side} and~\ref{subsec:independent-side}, we establish the desired
equivalence by expressing both choice probabilities as sums over
subsets of unavailable alternatives and matching their corresponding
terms.

\subsection{A summation identity} \label{subsec:summation-identity}

As shown later on, paired normalizers of the form $\psi_s$ and $\psi_{N-s-1}$ arise when a
ranking is divided into the alternatives before and after a specified
position. Our first step consists of expressing a sum involving the logistic
cumulative distribution function in terms of these normalizers, forming the algebraic
basis of our representation proof. To arrive at this identity, for $0\le s\le N-1$, let
\[ D_N(t) ~~=~~ \prod_{\ell=0}^{N}(1+tq^\ell) \qquad \text{and} \qquad  T_{N,s} ~~=~ \sum_{k\in\bbZ}\frac{q^{k(s+1)}}{D_N(q^k)}. \]
In what follows, we proceed by proving the next claim.

\begin{lemma}\label{lem:sum}
For every pair of integers $N\ge1$ and $0 \leq s \leq N - 1$, and for every $q\in(0,1)$,
\[
(1-q)\cdot T_{N,s}
~~=~~
q^{-s(s+1)/2}\cdot
\frac{\psi_s\psi_{N-s-1}}{\psi_N}.
\]
\end{lemma}

\paragraph{Convergence.} We first verify that the infinite summation defining $T_{N,s}$ indeed converges, so that subsequent manipulations of
such sums are well-defined. To this end, for every integer $K\ge1$, note that since $D_N(q^k)\ge1$, the positive tail of $T_{N,s}$ satisfies
\[
 \sum_{k=K}^{\infty}
 \frac{q^{k(s+1)}}{D_N(q^k)}
 ~~\le~~
 \sum_{k=K}^{\infty}q^{k(s+1)}
 ~~=~~
 \frac{q^{K(s+1)}}{1-q^{s+1}}.
\]
For its negative tail, we have $D_N(q^k) \ge  \prod_{\ell=0}^{N}q^{k+\ell} =
 q^{(N+1)k+N(N+1)/2}$, and consequently, 
\begin{align*}
 \sum_{k=-\infty}^{-K}
 \frac{q^{k(s+1)}}{D_N(q^k)}
 ~~&\le ~~
 q^{-N(N+1)/2} \cdot 
 \sum_{k=-\infty}^{-K}q^{k(s-N)}\\
 &=~~
 q^{-N(N+1)/2}  \cdot 
 \sum_{j=K}^{\infty}q^{j(N-s)}\\
 &=~~
 q^{-N(N+1)/2} \cdot 
 \frac{q^{K(N-s)}}{1-q^{N-s}}.
\end{align*}
Since $0<q<1$ and $0\le s\le N-1$, both bounds are finite
and tend to zero as $K$ tends to infinity. Thus, the series defining
$T_{N,s}$ converges absolutely.

% \paragraph{Base case: $N=1$.}
% Our proof proceeds by induction on $N$. When $N=1$, we must
% have $s=0$. In this case, $D_1(q^k)=(1+q^k)(1+q^{k+1})$, and therefore,
% \begin{align*}
%  T_{1,0}
%  ~~&=~~
%  \lim_{K\to\infty}
%  \sum_{k=-K}^{K}
%  \frac{q^k}{(1+q^k)(1+q^{k+1})}\\
%  &=~~
%  \frac{1}{1-q} \cdot 
%  \lim_{K\to\infty}
%  \sum_{k=-K}^{K}
%  \left(
%  \frac{1}{1+q^{k+1}}-\frac{1}{1+q^k}
%  \right)\\
%  &=~~
%  \frac{1}{1-q} \cdot 
%  \lim_{K\to\infty}
%  \left(
%  \frac{1}{1+q^{K+1}}-\frac{1}{1+q^{-K}}
%  \right)\\
%  &=~~
%  \frac{1}{1-q},
% \end{align*}
% where the third equality follows by telescoping.
% Since $\psi_0=\psi_1=1$, we have just proved the claimed identity for $N=1$.

\paragraph{Derivation.} 
We begin by presenting the next auxiliary claim, whose proof is provided in Appendix~\ref{app:proof_clm_sum-successive}. This claim can be interpreted as a recurrence that determines every $T_{N,s}$ from
the starting value $T_{N,0}$, thereby reducing our proof to evaluating that initial term.

\begin{claim} \label{clm:sum-successive}
For every $N \geq 2$ and $0\le s\le N-2$,
\[ T_{N,s+1}
 ~~=~~
 q^{-(s+1)} \cdot 
 \frac{1-q^{s+1}}{1-q^{N-s-1}} \cdot T_{N,s}. \]
\end{claim}

To employ this connection, we next determine $T_{N,0}$ by observing that for every $N \geq 1$,
\begin{align}
T_{N,0}
~~&=~~
\frac{1}{1-q^N}\cdot
\lim_{K\to\infty}
\sum_{k=-K}^{K}
\left(
\frac{1}{D_{N-1}(q^{k+1})}
-
\frac{1}{D_{N-1}(q^k)}
\right) \label{eqn:initial_sum_a1}\\
&=~~
\frac{1}{1-q^N}\cdot
\lim_{K\to\infty}
\left(
\frac{1}{D_{N-1}(q^{K+1})}
-
\frac{1}{D_{N-1}(q^{-K})}
\right) \label{eqn:initial_sum_a2}\\
&=~~
\frac{1}{1-q^N}. \label{eqn:initial_sum_a3}
\end{align}
Here, equality~\eqref{eqn:initial_sum_a1} holds since 
$D_N(t)=(1+t) \cdot D_{N-1}(qt)=(1+tq^N) \cdot D_{N-1}(t)$, evaluated at $t = q^k$. Equality~\eqref{eqn:initial_sum_a2} follows by telescoping. Equality~\eqref{eqn:initial_sum_a3} is obtained by noting that $\lim_{K\to\infty} D_{N-1}(q^{K+1}) = 1$ and $\lim_{K\to\infty} D_{N-1}(q^{-K}) = \infty$.

To complete our proof, let us fix any $0\le s\le N-1$. Then, by successively applying Claim~\ref{clm:sum-successive}, we conclude that 
\begin{align}
(1-q)\cdot T_{N,s}
~~&=~~
(1-q)\cdot T_{N,0}\cdot
\prod_{j=1}^{s}
\left(
q^{-j}\cdot\frac{1-q^j}{1-q^{N-j}}
\right)
\label{eqn:general_sum_1}\\
&=~~
q^{-s(s+1)/2}\cdot
\frac{1-q}{1-q^N}\cdot
\prod_{j=1}^{s}\frac{1-q^j}{1-q^{N-j}}
\label{eqn:general_sum_2}\\
&=~~
q^{-s(s+1)/2}\cdot
\frac{1}{[N]_q}\cdot
\prod_{j=1}^{s}\frac{[j]_q}{[N-j]_q}
\label{eqn:general_sum_3}\\
&=~~
q^{-s(s+1)/2}\cdot
\frac{1}{[N]_q}\cdot
\frac{\psi_s\psi_{N-s-1}}{\psi_{N-1}}
\label{eqn:general_sum_4}\\
&=~~
q^{-s(s+1)/2}\cdot
\frac{\psi_s\psi_{N-s-1}}{\psi_N}.
\label{eqn:general_sum_5}
\end{align}
In this case, equality~\eqref{eqn:general_sum_1} follows by applying
Claim~\ref{clm:sum-successive} for successive indices
$0,\ldots,s-1$.
Equality~\eqref{eqn:general_sum_2} substitutes
$T_{N,0}= \frac{ 1 }{ 1-q^N }$, according to~\eqref{eqn:initial_sum_a3}.
Equality~\eqref{eqn:general_sum_3} follows from
$[m]_q= \frac{ 1-q^m }{ 1-q }$.
To obtain equality~\eqref{eqn:general_sum_4}, note that its  numerator and denominator products equal $\psi_s$ and $\frac{ \psi_{N-1} }{ \psi_{N-s-1} }$, respectively.
Finally, equality~\eqref{eqn:general_sum_5} holds since
$\psi_N=[N]_q \cdot \psi_{N-1}$.

\subsection{Decomposing the Mallows choice probability} \label{subsec:mallows-side}

To establish Theorem~\ref{thm:representation}, given a nonempty set $A\subseteq\UU^+$ and an alternative $i\in A$, our analysis will compare the choice probabilities $\prpar{U_i=\max_{j\in A}U_j}$ and $\pihat(i,A)$ by expressing each as
a sum indexed by subsets of unavailable alternatives. Under the
Mallows model, these subsets can be naturally interpreted as specifying which alternatives appear before $i$ in the sampled
ranking, and we therefore develop this direction first.

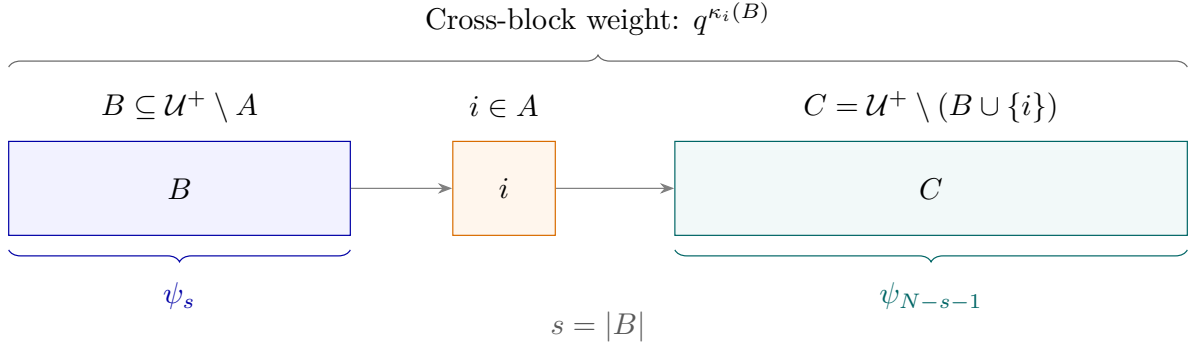
\begin{figure}[htpb!]
\centering
\resizebox{0.98\linewidth}{!}{%
\begin{tikzpicture}[
    font=\small, >=Stealth,
    brace/.style={decorate, decoration={brace,amplitude=4pt}},
    block/.style={draw, minimum height=11mm, inner sep=0pt}
]
% The ranking has three consecutive blocks.
\node[block, draw=blue!65!black, fill=blue!5,
      minimum width=40mm] (B) at (2.2,0) {$B$};
\node[block, draw=orange!85!black, fill=orange!7,
      minimum width=12mm] (i) at (6,0) {$i$};
\node[block, draw=teal!80!black, fill=teal!5,
      minimum width=60mm] (C) at (11,0) {$C$};
\draw[->, black!50] (B.east) -- (i.west);
\draw[->, black!50] (i.east) -- (C.west);

% All predecessors of the chosen alternative are unavailable.
\node at (2.2,0.96) {$B\subseteq\mathcal U^+\setminus A$};
\node at (6,0.96) {$i\in A$};
\node at (11,0.96) {$C=\mathcal U^+\setminus(B\cup\{i\})$};

% Cross-block inversions are fixed once B is specified.
\draw[brace, black!60] (0.2,1.42) -- (14,1.42)
    node[midway, above=7pt, text=black]
    {Cross-block weight: $q^{\kappa_i(B)}$};

% Summation over the internal orders of B and C.
\draw[brace, blue!65!black] (4.2,-0.72) -- (0.2,-0.72)
    node[midway, below=7pt] {$\psi_s$};
\draw[brace, teal!80!black] (14,-0.72) -- (8,-0.72)
    node[midway, below=7pt] {$\psi_{N-s-1}$};
\node[black!65] at (7.1,-1.62) {$s=|B|$};
\end{tikzpicture}%
}
\caption{For fixed $B$, inversions between blocks contribute
$q^{\kappa_i(B)}$, while summing over internal orders contributes
$\psi_s\psi_{N-s-1}$.}
\label{fig:block-decomposition}
\end{figure}

\paragraph{Partitioning by the predecessor set.} 
With respect to the Mallows model, alternative $i$ is selected from $A$ precisely when every alternative
appearing before it is unavailable. As such, this event can be partitioned into pairwise disjoint events
indexed by subsets $B\subseteq\UU^+\setminus A$, where the event
associated with $B$ consists of rankings in which the alternatives in $B$ are those preceding $i$. Letting $C=\UU^+\setminus(B\cup\{i\})$, each such ranking
consists of an ordering of $B$, followed by $i$, followed by an
ordering of $C$. Figure~\ref{fig:block-decomposition}
illustrates this decomposition and the corresponding
contributions to the Mallows weight.

\paragraph{Counting inversions between blocks.} Let us fix a subset $B\subseteq\UU^+\setminus A$, with $s=|B|$
standing for its size.
To compute the probability that precisely the alternatives in $B$
precede $i$ in the sampled ranking, let
$b_i(B)=|\{j\in B:\omega(j)<\omega(i)\}|$ be the number of
alternatives in $B$ that are ranked higher than $i$ in the
modal permutation $\omega$. With $i$ being fixed, the number of inversions between $B$, $\{ i \}$, and $C$ can be expressed as follows:
\begin{itemize}
    \item \textit{Between $B$ and $C \cup \{i\}$}: Letting $u_1<\cdots<u_s$ be the modal ranks of the alternatives in $B$, an alternative with rank $u_\ell$ forms an inversion with each of the $u_\ell-\ell$ alternatives outside $B$ that have smaller ranks. Thus, the total number of such inversions is $\sum_{\ell=1}^{s}(u_\ell-\ell)  =
 \sum_{j\in B}\omega(j)-\frac{s(s+1)}{2}$. 

 \item \textit{Between $\{i\}$ and $C$}:
Since $i$ precedes every alternative in $C$ in the sampled ranking,
an inversion occurs precisely for those alternatives in $C$
whose modal ranks are smaller than $\omega(i)$. There are $\omega(i)-1$ alternatives in $\mathcal U^+$
whose modal ranks are smaller than $\omega(i)$, of which $b_i(B)$ belong to $B$. Thus, the number of inversions between $i$ and $C$ is
$\omega(i)-1-b_i(B)$.    
\end{itemize}
Aggregating both contributions, we denote the total number of inversions between the three
blocks by
\begin{equation}\label{eq:cross-inversions}
\kappa_i(B)
~~=~~
\sum_{j\in B}\omega(j)-\frac{s(s+1)}{2}
+\omega(i)-1-b_i(B).
\end{equation}
This quantity depends only on $B$ and the fixed alternative $i$,
regardless of the internal orderings of $B$ and $C$.

\paragraph{Summing over internal orderings.} Once $B$ is fixed, each ordering of $B$ and each ordering of $C$
jointly determine exactly one ranking whose predecessor set for $i$ is $B$. Here, the inversions between the three blocks contribute the same
factor $q^{\kappa_i(B)}$ to the Mallows weight of every such
ranking. The remaining factors depend only on the inversions
within $B$ and within $C$. We can therefore separately sum over the internal orderings of these two
blocks. For $B$, this sum is the Mallows normalizer
$\psi_s$, computed with respect to the modal order restricted
to $B$; similarly, the analogous sum for $C$ is $\psi_{N-s-1}$.
Thus, the total weight of rankings with predecessor set $B$ is
$q^{\kappa_i(B)}\psi_s\psi_{N-s-1}$, and dividing by the normalizer $\psi_N$, the probability of this
event is $\frac{q^{\kappa_i(B)}\psi_s\psi_{N-s-1}}{\psi_N}$. Since these events are pairwise disjoint and together constitute the
event that $i$ is selected from $A$, by summing their probabilities
over all $B\subseteq\UU^+\setminus A$, we have 
\begin{equation}\label{eq:expanded}
\pihat(i,A)
~~=~~
\frac{1}{\psi_N}\cdot
\sum_{B\subseteq\UU^+\setminus A}
q^{\kappa_i(B)}
\psi_{|B|}\psi_{N-|B|-1}.
\end{equation}

\subsection{Decomposing the independent-utility choice probability}
\label{subsec:independent-side}

We next show that, given a nonempty set $A\subseteq\UU^+$
and an alternative $i\in A$, the choice probability $\prpar{U_i=\max_{j\in A}U_j}$ of this alternative according to the independent utilities
$\{ U_j \}_{j \in \UU^+}$ matches expression~\eqref{eq:expanded} to the letter. For this purpose, let $\mathcal{M} =\UU^+\setminus A$ be
the set of unavailable alternatives. Our analysis first conditions on
the score $X_i$ of $i$ and then expands the resulting expression into
terms indexed by subsets of $\mathcal{M}$.

\paragraph{Conditioning on the alternative's score.}
For any integer $k$, conditional on $X_i=k$, alternative $i$
is selected from $A$ precisely when every available competitor
preceding it in the modal ranking has score strictly below $k$,
and every available competitor following it has score at most $k$.
Given  relationship~\eqref{eq:real-utility} between $\{U_j\}_{j\in A}$ and
$\{X_j\}_{j\in A}$, together with the independence of these scores, we infer that 
\begin{equation}\label{eq:ind-choice}
\pr{U_i=\max_{j\in A}U_j}
~~=~~
\sum_{k\in\bbZ}
\pr{ X_i = k }\cdot
\prod_{\substack{j\in A:\\\omega(j)<\omega(i)}}F_{X_j}(k-1)\cdot
\prod_{\substack{j\in A:\\\omega(j)>\omega(i)}}F_{X_j}(k).
\end{equation}
According to the cumulative distribution function~\eqref{eqn:CDF_Xi} of our scores, we know that $F_{X_j}(k)= \frac{ 1 }{ 1+q^{k+\omega(j)} }$. Thus, with the shorthand notation $t=q^k$, the probability mass of $X_i$ can be expressed as
\[ \pr{ X_i = k } ~~=~~
F_{X_i}(k)-F_{X_i}(k-1)
~~=~~
\frac{(1-q)tq^{\omega(i)-1}}
{(1+tq^{\omega(i)-1})(1+tq^{\omega(i)})}.
\]

\paragraph{Completing the denominator.}
We now consider the denominator of the entire $k$-th summand
in representation~\eqref{eq:ind-choice}, after substituting the explicit formulas
for $\pr{X_i=k}$ and the competing-score cumulative distribution functions. Here, $\pr{X_i=k}$ contributes the two factors
$(1+tq^{\omega(i)-1})(1+tq^{\omega(i)})$, whereas each competitor $j \in A \setminus \{ i \}$ contributes one additional factor:
$1+tq^{\omega(j)-1}$ when $\omega(j)<\omega(i)$,
and $1+tq^{\omega(j)}$ when $\omega(j)>\omega(i)$.  It is worth noting that a competitor preceding $i$ contributes an exponent
$\omega(j)-1\le\omega(i)-2$, whereas a competitor following $i$
contributes an exponent $\omega(j)\ge\omega(i)+1$.
Thus, neither of the exponents $\omega(i)-1$ and $\omega(i)$
contributed by $i$ appears among the competitors' factors. As a result, when all alternatives are available, these factors collectively
include every $1+tq^\ell$, for $0\le\ell\le N$, exactly once.
Thus, in this simpler case, the denominator of the full summand is
 $D_N(t) = \prod_{\ell=0}^{N}(1+tq^\ell)$. For a general availability  set $A$, the missing factors correspond
to the unavailable alternatives $j\in \mathcal{M}$, where the factor associated with each such alternative is
$1+tq^{\omega(j)-\ind[\omega(j)<\omega(i)]}$. Therefore,
\begin{equation}\label{eq:completed-denominator}
\pr{U_i=\max_{j\in A}U_j}
~~=~~
(1-q)q^{\omega(i)-1}\cdot
\sum_{k\in\bbZ}
\frac{q^k}{D_N(q^k)}\cdot
\prod_{j\in \mathcal{M}}
\left(1+q^{k+\omega(j)-\ind[\omega(j)<\omega(i)]}\right).
\end{equation}

\paragraph{Expanding over unavailable alternatives.} Now, expanding $\prod_{j\in \mathcal{M}}
(1+q^{k+\omega(j)-\ind[\omega(j)<\omega(i)]})$
amounts to choosing either the constant term $1$ or the nonconstant
term $q^{k+\omega(j)-\ind[\omega(j)<\omega(i)]}$ from each factor, and we make use of $B\subseteq \mathcal{M}$ to index the
alternatives for which the nonconstant term is chosen.
Recalling that $b_i(B)=|\{j\in B:\omega(j)<\omega(i)\}|$, it follows that this expansion can be written as 
\begin{equation} \label{eqn:expansion_product}
\prod_{j\in \mathcal{M}}
\left(1+q^kq^{\omega(j)-\ind[\omega(j)<\omega(i)]}\right)
~~=~~
\sum_{B\subseteq \mathcal{M}}
q^{k|B|+\sum_{j\in B}\omega(j)-b_i(B)}.    
\end{equation}
Consequently, we conclude the proof by noticing that
\begin{align}
\pr{U_i=\max_{j\in A}U_j}
~~&=~~
(1-q)q^{\omega(i)-1}\cdot
\sum_{B\subseteq\mathcal{M}}
q^{\sum_{j\in B}\omega(j)-b_i(B)}\cdot T_{N,|B|}
\label{eqn:independent_expansion_1}\\
&=~~
\frac{1}{\psi_N}\cdot
\sum_{B\subseteq\mathcal{M}}
q^{\omega(i)-1+\sum_{j\in B}\omega(j)-b_i(B)-|B|(|B|+1)/2}
\psi_{|B|}\psi_{N-|B|-1}
\label{eqn:independent_expansion_2}\\
&=~~
\frac{1}{\psi_N}\cdot
\sum_{B\subseteq\mathcal{M}}
q^{\kappa_i(B)}
\psi_{|B|}\psi_{N-|B|-1}
\label{eqn:independent_expansion_3}\\
&=~~ \pihat(i,A).
\label{eqn:independent_expansion_4}
\end{align}
Here, equality~\eqref{eqn:independent_expansion_1} follows by
substituting expansion~\eqref{eqn:expansion_product} into
equation~\eqref{eq:completed-denominator} and interchanging the finite sum over $B$ with the convergent sum over $k$. For each fixed subset $B$, the resulting inner sum is $\sum_{k\in\bbZ}
\frac{q^{k(|B|+1)}}{D_N(q^k)} = T_{N,|B|}$.
Equality~\eqref{eqn:independent_expansion_2} applies
Lemma~\ref{lem:sum} separately to each summand, with $s=|B|$.
This application is valid since $i\in A$ and
$B\subseteq\mathcal{M}=\UU^+\setminus A$, ensuring that
$0\le|B|\le N-1$.
Equality~\eqref{eqn:independent_expansion_3} identifies the
exponent of $q$ with $\kappa_i(B)$, as defined
in~\eqref{eq:cross-inversions}.
Finally, equality~\eqref{eqn:independent_expansion_4} is exactly the Mallows choice probability expansion
in~\eqref{eq:expanded}, recalling that
$\mathcal{M}=\UU^+\setminus A$.
\section{Revenue Bounds for Finite-Support Reduction}\label{sec:finite}

In what follows, we establish the expected revenue bounds stated in Lemmas~\ref{lem:coarse} and~\ref{lem:clip}, utilized to argue that  our finite-support
reduction indeed produces near-optimal assortments.
In Section~\ref{subsec:proof_lem_coarse}, we analyze  grouping effects by conditioning on the
competing scores and examining the extent to which the
thresholds required for an alternative to win can be displaced. This quantity is then translated
into a multiplicative bound on expected revenue. Subsequently, Section~\ref{subsec:proof_lem_clip} studies truncation by coupling the grouped
and truncated scores, to bound the probability that any score
changes, yielding a negligible additive revenue loss
for every assortment.

\subsection{Post-grouping revenue: Proof of Lemma~\ref{lem:coarse}} \label{subsec:proof_lem_coarse}

\paragraph{The sufficient inequality.} To show that $e^{-\eps} \cdot R_X(S)\le R_Y(S)\le e^\eps \cdot  R_X(S)$ for every assortment $S\subseteq\UU$, it suffices to argue that for every set $A \subseteq \UU^+$ and alternative $i \in A$, we have
\begin{equation} \label{eq:coarse}
 q^{\Delta-1} \cdot \pihat_X(i,A) ~~\le~~ \pihat_Y(i,A)
 ~~\le~~ q^{-(\Delta-1)} \cdot \pihat_X(i,A),
\end{equation}
where $\pihat_X(i,A)$ and $\pihat_Y(i,A)$ are the probabilities
that alternative $i$ is selected from $A$ under the scores
$\{X_j\}_{j\in A}$ and $\{Y_j\}_{j\in A}$, respectively. Indeed, specifically for $A=S^+$,
we obtain
\begin{align*}
 e^{-\eps} \cdot R_X(S)
 ~~&\le~~ q^{\Delta-1} \cdot R_X(S)\\
 &=~~ \sum_{i\in S}r_i q^{\Delta-1} \cdot \pihat_X(i,S^+)\\
 &\le~~ \sum_{i\in S}r_i \cdot \pihat_Y(i,S^+)\\
 &=~~ R_Y(S)\\
 &\le~~ \sum_{i\in S}r_i q^{-(\Delta-1)} \cdot\pihat_X(i,S^+)\\
 &=~~ q^{-(\Delta-1)} \cdot R_X(S)\\
 &\le~~ e^\eps \cdot R_X(S),
\end{align*}
where the first and last inequalities hold since  $q=e^{-\theta}$ and
$\theta(\Delta-1)\le\eps$, implying that $q^{\Delta-1}\ge e^{-\eps}$ and $q^{-(\Delta-1)}\le e^\eps$.

To derive inequality~\eqref{eq:coarse}, we utilize independence to fix all competing scores and examine the remaining score
through a single threshold. As it turns out, our grouping operation moves this threshold
by an appropriately bounded amount, and we will explain how this displacement can be converted into the desired relation between $\pihat_X(i,A)$ and $\pihat_Y(i,A)$. Given that inequality~\eqref{eq:coarse} is immediate when $A=\{i\}$, we assume that $A$ contains
at least one additional alternative. 

\begin{figure}[htbp!]
\centering
\resizebox{0.98\linewidth}{!}{%
\begin{tikzpicture}[font=\small, >=Stealth]
\node[anchor=west, blue!65!black] at (8.5,5.55)
    {$\tau_1$: original};
\node[anchor=west, orange!85!black] at (11.8,5.55)
    {$\tau_2$: grouped};

% Panel (a): j has priority; panel (b): i has priority.
\foreach \y/\panel/\priority/\original/\grouped in {
    3.35/a/{\omega(j)<\omega(i)}/7/8,
    0.35/b/{\omega(i)<\omega(j)}/6/4} {
    \node[anchor=west] at (0,{\y+1.45})
        {(\panel) $\priority$};
    \node[anchor=west] at (0,{\y+0.25}) {$X_i$};

    % Grouping intervals.
    \foreach \g in {0,1,2} {
        \draw[draw=black!20, fill=blue!4]
            ({2.65+3.92*\g},\y)
            rectangle ({6.29+3.92*\g},{\y+0.72});
        \node[black!55]
            at ({4.47+3.92*\g},{\y+0.38}) {$Y_i=\g$};
    }

    % Integer scores.
    \foreach \k in {0,...,11} {
        \fill[black!50] ({3+0.98*\k},\y) circle (0.9pt);
        \node at ({3+0.98*\k},{\y-0.36}) {$\k$};
    }

    % Original and grouped thresholds.
    \draw[thick, blue!65!black]
        ({3+0.98*\original},{\y-0.07}) --
        ({3+0.98*\original},{\y+1.10});
    \draw[thick, orange!85!black]
        ({3+0.98*\grouped},{\y-0.07}) --
        ({3+0.98*\grouped},{\y+1.10});

    \node[blue!65!black]
        at ({3+0.98*\original},{\y+1.36}) {$\tau_1$};
    \node[orange!85!black]
        at ({3+0.98*\grouped},{\y+1.36}) {$\tau_2$};

    \draw[->, thick, orange!85!black]
        ({3+0.98*\original},{\y+0.88}) --
        ({3+0.98*\grouped},{\y+0.88});
}
%\node at (7.7,-0.8) {$|\tau_1-\tau_2|\le h-1$};
\end{tikzpicture}%
}
\caption{With $\Delta=4$ and a single competitor having $X_j=6$, grouping can
raise or lower the winning threshold, depending on
which alternative has priority in a tie.}
\label{fig:grouping-thresholds}
\end{figure}
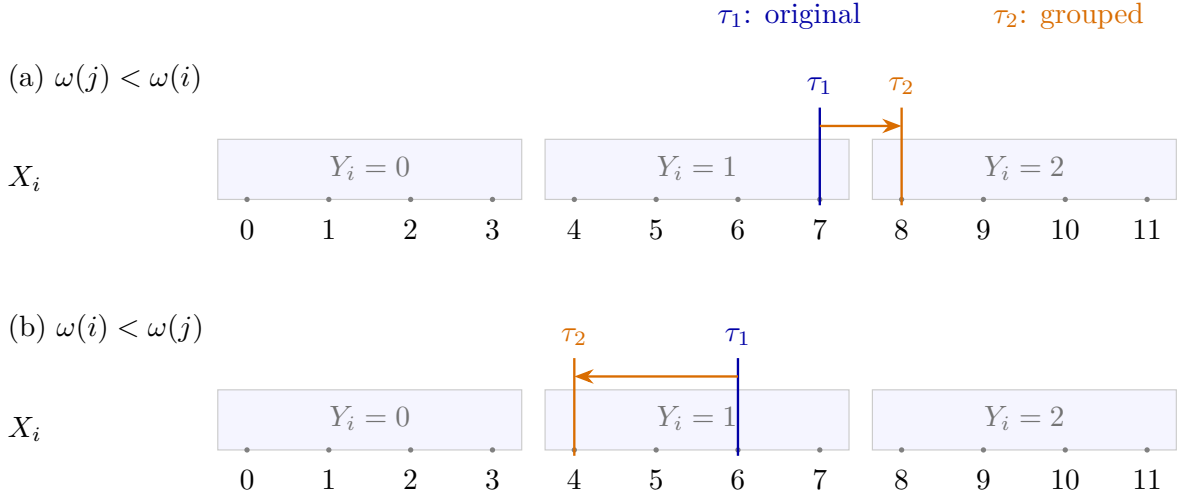

\paragraph{Step 1: Conditional winning thresholds.} For any realization $X_{-i} = x_{-i}$ of all competitors' scores, alternative $i$ wins under the original scores
$\{X_j\}_{j\in A}$ if and only if 
\[
X_i ~~\geq~~ \underbrace{ \max_{j\in A\setminus\{i\}}
 \left\{ x_j+\ind[\omega(j)<\omega(i)]\right\} }_{ \tau_1 }.
\]
Under the same realization of $X_{-i}$, each competing
alternative $j$ has grouped score $Y_j=\lfloor \frac{ x_j }{ \Delta }\rfloor$.
Thus, alternative $i$ wins
under $\{Y_j\}_{j\in A}$ if and only if $ Y_i
 \ge
 \max_{j\in A\setminus\{i\}}
\{ 
 \lfloor\frac{x_j}{\Delta}\rfloor
 +\ind[\omega(j)<\omega(i)]
 \}$. Since the threshold on the right is an integer, this condition is equivalent to
\[
 X_i ~~\geq~~ \underbrace{
 \max_{j\in A\setminus\{i\}}
 \Delta \cdot \left\{
 \left\lfloor\frac{x_j}{\Delta}\right\rfloor
 +\ind[\omega(j)<\omega(i)]
 \right\} }_{ \tau_2 }.
\]
Figure~\ref{fig:grouping-thresholds} illustrates that grouping can
either raise or lower the winning threshold, depending on modal
priority. We next show that the magnitude of this change is 
at most $\Delta-1$.

To compare these thresholds, note that the terms corresponding to each competitor $j$ in the  thresholds above differ by
\begin{align*}
 &\Delta \cdot \left(
 \left\lfloor\frac{x_j}{\Delta}\right\rfloor
 +\ind[\omega(j)<\omega(i)]
 \right)
 -
 \left(x_j+\ind[\omega(j)<\omega(i)]\right)\\
 &\hspace{2cm}
 =~~
 (\Delta-1) \cdot \ind[\omega(j)<\omega(i)]-
 \left( x_j - \Delta \cdot \left\lfloor\frac{x_j}{\Delta}\right\rfloor \right) \\
 &\hspace{2cm}
 \in~~ [-(\Delta-1),\Delta-1].
 \end{align*}
Since corresponding terms differ by at most $\Delta-1$ in
absolute value, their maxima satisfy $|\tau_2-\tau_1|\leq\Delta-1$.

\paragraph{Step 2: Converting threshold displacement to inequality~\eqref{eq:coarse}.} We now translate this bound into
a comparison of the conditional winning probabilities. According to~\eqref{eqn:CDF_Xi}, we know that $F_{X_i}(k) = \frac{1}{1+q^{k+\omega(i)}}$ for every $k \in \bbZ$, implying that $\prpar{X_i\ge k} =\frac{q^{k+\omega(i)-1}}{1+q^{k+\omega(i)-1}}$. By independence, conditioning on the competing scores $X_{-i}$ does not
change the distribution of $X_i$. Thus, the conditional winning
probabilities under the original and grouped scores are $\prpar{X_i\ge \tau_1}$ and $\prpar{X_i\ge \tau_2}$, respectively. We proceed by considering two cases, depending on the order between $\tau_1$ and $\tau_2$:
\begin{itemize}
    \item \textit{When $\tau_2\ge\tau_1$}: In this scenario, based on the tail formula above, we have
    \[ \frac{\pr{X_i\ge\tau_2}}{\pr{X_i\ge\tau_1}}
    ~~=~~
    q^{\tau_2-\tau_1}\cdot
    \frac{1+q^{\tau_1+\omega(i)-1}}
         {1+q^{\tau_2+\omega(i)-1}}
    ~~\ge~~ q^{\tau_2-\tau_1}
    ~~\ge~~ q^{\Delta-1},
    \]
    where the last inequality holds since $0\le \tau_2-\tau_1\le\Delta-1$.
    In the other direction, 
    \[
    \pr{X_i\ge\tau_2}
    ~~\le~~ \pr{X_i\ge\tau_1}
    ~~\le~~ q^{-(\Delta-1)}\cdot\pr{X_i\ge\tau_1}.
    \]

    \item \textit{When $\tau_2<\tau_1$}: Here, $0<\tau_1-\tau_2\le\Delta-1$, and a similar argument leads to
    \[ \frac{\pr{X_i\ge\tau_2}}{\pr{X_i\ge\tau_1}}
    ~~=~~
    q^{-(\tau_1-\tau_2)}\cdot
    \frac{1+q^{\tau_1+\omega(i)-1}}
         {1+q^{\tau_2+\omega(i)-1}}
    ~~\le~~ q^{-(\tau_1-\tau_2)}
    ~~\le~~ q^{-(\Delta-1)}. \]
    In the other direction, 
    \[
    \pr{X_i\ge\tau_2}
    ~~\ge~~ \pr{X_i\ge\tau_1}
    ~~\ge~~ q^{\Delta-1}\cdot\pr{X_i\ge\tau_1}.
    \]
\end{itemize}
Thus, in both cases, the conditional winning probabilities satisfy
\[
q^{\Delta-1}\cdot\pr{X_i\ge\tau_1}
~~\le~~ \pr{X_i\ge\tau_2}
~~\le~~ q^{-(\Delta-1)}\cdot\pr{X_i\ge\tau_1}.
\]
Since these bounds hold for every realization of the competing
scores $X_{-i}$, averaging over these realizations yields
inequality~\eqref{eq:coarse}, as $\pihat_X(i,A)
= \exsubpar{X_{-i}} { \prpar{X_i\ge\tau_1 | X_{-i}} }$ and $\pihat_Y(i,A)
= \exsubpar{X_{-i}} { \prpar{X_i\ge\tau_2 | X_{-i}} }$.

\subsection{Post-truncation revenue: Proof of Lemma~\ref{lem:clip}} \label{subsec:proof_lem_clip}

We begin by recalling that, for any alternative $i\in\UU^+$, its grouped and truncated scores are related via $\bar Y_i=\max\{-\Gamma,\min\{Y_i,\Gamma\}\}$. Thus, $Y_i$ and $\bar Y_i$ differ precisely when
$Y_i\notin[-\Gamma,\Gamma]$. Let $\mathcal E$ be the event that at least one such difference exists, i.e., $\mathcal{E} = \bigcup_{i\in\UU^+}\{Y_i\ne\bar Y_i\}$. The crux of our proof resides in arguing that $\prpar{Y_i\ne\bar Y_i}
\le \frac{2 \eps q^{N-1} }{ N^2 }$ for all $i\in\UU^+$. We first explain why this claim suffices to attain the desired expected revenue bound, and then turn our attention to its proof.

\paragraph{Effect on expected revenue.}
Given an assortment $S\subseteq\UU$, let $I_Y(S)$ and
$I_{\bar Y}(S)$ be the random alternatives selected from $S^+$
under the grouped and truncated scores, respectively.
On the event $\{ I_Y(S) = I_{\bar Y}(S) \}$, both models select the same alternative, and their realized revenues are  identical. On the complementary event, both realized revenues lie in $[0,r_{\max}]$,
implying that their absolute difference is at most $r_{\max}$. Therefore, 
\begin{align}
 |R_{\bar Y}(S)-R_Y(S)|
 ~~&=~~ \left|
 \ex{r_{I_{\bar Y}(S)} } - \ex{ r_{I_Y(S)}}
 \right| \nonumber \\
~~ &\le~~ \pr{ I_Y(S) \neq I_{\bar Y}(S) } \cdot r_{\max} \nonumber \\
 &\le~~ \pr{\mathcal E} \cdot r_{\max} \label{eqn:lem_clip_1} \\
 &\le~~ 2\eps \cdot \frac{ q^{N-1} }{ N } \cdot r_{\max} \label{eqn:lem_clip_2} \\
 & \leq~~2\eps \cdot \opt. \label{eqn:lem_clip_3}
\end{align}
Here, inequality~\eqref{eqn:lem_clip_1} holds since $\{ I_Y(S) \neq I_{\bar Y}(S) \} \subseteq \mathcal{E}$. Inequality~\eqref{eqn:lem_clip_2} follows by applying the union bound with $\prpar{Y_i\ne\bar Y_i}
\le \frac{ 2\eps q^{N-1} }{ N^2 }$ for each alternative $i\in\UU^+$. Lastly, inequality~\eqref{eqn:lem_clip_3} holds since $\opt \geq \frac{ q^{N-1} }{ N } \cdot r_{\max}$, by Claim~\ref{clm:LB_B_OPT}.

\paragraph{Proving $\prpar{Y_i\ne\bar Y_i}
\le \frac{ 2\eps q^{N-1} }{ N^2 }$.} As previously mentioned, $Y_i$ and $\bar Y_i$ differ precisely when
$Y_i\notin[-\Gamma,\Gamma]$. To evaluate the probability of this event, note that since $Y_i=\lfloor \frac{ X_i }{ \Delta }\rfloor$
and $X_i$ is integer-valued, 
\[ \pr{Y_i<-\Gamma} ~~=~~
F_{X_i}(-\Delta\Gamma-1)
~~=~~ \frac{1}{1+q^{-\Delta\Gamma-1+\omega(i)}} ~~ \le~~ q^{\Delta\Gamma+1-\omega(i)}
~~\le~~ q^{\Delta\Gamma+1-N}, \]
where the second equality holds since $F_{X_i}(k) =\frac{1}{1+q^{k+\omega(i)}}$, as shown in equation~\eqref{eqn:CDF_Xi}. Similarly,
\[
\pr{Y_i>\Gamma}
~~=~~ 1-F_{X_i}\bigl(\Delta(\Gamma+1)-1\bigr)
~~=~~ \frac{q^{\Delta(\Gamma+1)+\omega(i)-1}}
           {1+q^{\Delta(\Gamma+1)+\omega(i)-1}}
~~\le~~ q^{\Delta\Gamma}
~~\le~~ q^{\Delta\Gamma+1-N}.
\]
Combining these two tail bounds, we obtain
\begin{align}
\pr{Y_i\ne\bar Y_i}
~~&=~~ \pr{Y_i<-\Gamma}+\pr{Y_i>\Gamma} \nonumber\\
~~&\le~~ 2q^{\Delta\Gamma+1-N} \nonumber \\
~~&=~~ 2q^{1-N}e^{-\theta \Delta\Gamma}
\nonumber \\
~~&\le~~ 2q^{N-1}e^{-N^2/\eps}
\label{eqn:clip_probability_2}\\
~~&\le~~ \frac{2\eps q^{N-1}}{N^2}.
\label{eqn:clip_probability_3}
\end{align}
Here, to derive
inequality~\eqref{eqn:clip_probability_2}, note that $\theta \Delta\Gamma
\geq
\theta \Delta \cdot (2(N-1)+\frac{N^2}{\eps^2})
\geq
2(N-1)\theta+\frac{N^2}{\eps}$, since $\Gamma=\lceil \frac{ 2N^2 }{ \eps^2 }\rceil$, 
$\frac{ N^2 }{ \eps^2 } \geq 2(N-1)$, $\Delta\geq1$, and $\theta \Delta>\eps$. Inequality~\eqref{eqn:clip_probability_3} follows from $\frac{N^2}{\eps} \geq \ln(\frac{N^2}{\eps})$, which is an application of $\ln x\leq x$ for $x > 0$.
\section{Concluding Remarks}

Our approximation scheme rests on a fundamental distinction: Reproducing choice
probabilities can be easier than reproducing their underlying distribution over rankings. Motivated by this insight, we conclude  by describing an elegant probabilistic interpretation that originally guided our construction of the independent scores.
We then ask how far this approach can reach, both beyond classical Mallows
choice and beyond a single customer segment.

\paragraph{Interpretation through exponential waiting times.} Interestingly, the score distributions of  Section~\ref{subsec:milestone_1} admit a parallel derivation method, where each alternative draws two independent exponential waiting
times, and the logarithm of their ratio determines its noise.
Specifically, letting $E_i$ and $V_i$ be independent exponential
random variables of unit rate, with independent pairs for different
alternatives, suppose we define
$Q_i=\lceil \frac{ \ln(V_i/E_i) }{ \theta} \rceil$. Then, for every $k\in\bbZ$, the event
$\{Q_i\le k\}$ is equivalent to $\{E_i\ge q^k V_i\}$, since $q=e^{-\theta}$. Therefore, by conditioning on $V_i$,
\[
 \pr{Q_i\le k}
 ~~=~~
 \int_0^\infty \pr{E_i\ge q^k v} \cdot  e^{-v} \mathrm{d}v
 ~~=~~
 \int_0^\infty e^{-(1+q^k)v} \mathrm{d}v
 ~~=~~
 \frac{1}{1+q^k}.
\]
Thus, $Q_i$ has precisely the common noise distribution proposed
in equation~\eqref{eq:noise-cdf}, and setting $X_i=Q_i-\omega(i)$ recovers our
independent scores. This construction provides a direct way to sample $\{ X_i \}_{i \in \UU^+}$ from elementary random variables. Still, the integer-score formulation of Section~\ref{subsec:milestone_1} makes the winning thresholds and tail
bounds more explicit, better fitting our overall analysis.

\paragraph{Beyond the classical Mallows model.}
Our representation theorem raises a broader structural question: When can the
dependence in a ranking model be dispensed with at the level of choices?
Natural candidates include the generalized Mallows model with
position-dependent concentration parameters \citep{fligner1986}, which
retains a factorization through inversion coordinates, as well as Mallows
models for top-$k$ lists and their generalizations
\citep{chierichetti2018,haddadan2025}.
For these families, do realistic restrictions on the model parameters and
preference structure permit a single collection of independent utilities
to reproduce the choice probabilities simultaneously for every offered
set? For incomplete rankings, the answer could also depend on the rule
governing choices among unranked or incomparable alternatives.
To translate such a representation into an approximation scheme, one should concurrently construct a finite-support approximation of
polynomial size, with revenue losses controlled relative to the optimum.
The challenge is therefore twofold: Uncover the underlying independent utilities and
make their distributions computationally manageable.

\paragraph{Mixtures of Mallows segments.}
A different obstacle emerges when customer choice is governed by multiple Mallows
segments. For a fixed number $K$ of segments, can $(1-\epsilon)$-approximate assortments be computed in
$O(n^{f(K,1/\epsilon)})$ time, without any restrictions on product prices,
concentration parameters, or structural attributes of the modal rankings? Clearly, when our representation is separately applied within each segment, its resulting utilities are independent conditional on the customer's segment.
However, averaging over this shared random segment generally introduces
dependence across alternatives, preventing a direct application of 
independent-utility approximation schemes. The core difficulty lies in computing a single assortment that performs well across
these conditional representations. Toward this objective, a compelling starting point could be two Mallows segments, noting that known hardness results   do not establish any form of intractability for a fixed number of segments (see Section~\ref{sec:prior}).
Resolving this setting would clarify whether independence within customer
segments can still support a polynomial-time approximation scheme.

% --- Bibliography ---
\phantomsection
\addcontentsline{toc}{section}{References}
\bibliographystyle{plainnat}
\bibliography{BIB-Mallows}

% --- Appendix ---
\addtocontents{toc}{\protect\setcounter{tocdepth}{2}} % Stop TOC listing for Appendix
\appendix
\section{Additional Proofs}

\subsection{Proof of Claim~\ref{clm:LB_B_OPT}} \label{app:proof_clm_LB_B_OPT}

Let us focus on a single alternative $i$, appearing in position $\omega(i)$ of the modal permutation.
In the repeated insertion procedure (see Section~\ref{subsec:proof_thm_nonrum}), this alternative finishes first in the Mallows ranking if and only
if it is inserted in position $1$ and all subsequent alternatives are inserted in position $2$ or later. To account for this event, note that alternative $i$ is considered
at step $\omega(i)$, when its probability of being inserted first is
$\frac{ q^{\omega(i)-1} }{ [\omega(i)]_q }$. Then, at each subsequent step $s>\omega(i)$, the probability of avoiding
the first position is $1-\frac{q^{s-1}}{[s]_q} = \frac{[s-1]_q}{[s]_q}$. Since insertion positions are independently sampled, we have
\[
 \pr{\sigma(i)=1}
 ~~=~~
 \frac{q^{\omega(i)-1}}{[\omega(i)]_q} \cdot
 \prod_{s=\omega(i)+1}^{N}\frac{[s-1]_q}{[s]_q}
 ~~=~~
 \frac{q^{\omega(i)-1}}{[N]_q}
 ~~\ge~~
 \frac{q^{N-1}}{N},
\]
where the sole inequality above holds since  $\omega(i)\le N$,  $0<q\le1$, and $[N]_q=\sum_{\ell=0}^{N-1}q^\ell \leq N$. In particular, when offered alone, a maximum-price product is purchased
whenever it finishes first, meaning that the expected revenue
of this singleton assortment is at least $\frac{ q^{N-1} }{ N } \cdot r_{\max}$, providing the required lower bound on $\opt$. 

\subsection{Proof of Claim~\ref{clm:sum-successive}} \label{app:proof_clm_sum-successive}

First, it is important to point out that, for fixed $N\ge2$ and $0\le s\le N-2$, the series defining
$T_{N-1,s}$, $T_{N,s}$, and $T_{N,s+1}$ all converge
absolutely, justifying the splitting and reindexing below. 
We begin by deriving two useful relations. By noticing that  $D_N(t)=(1+tq^N) \cdot D_{N-1}(t)$, it follows that  
\begin{equation} \label{eq:sum-recursion-one}
 T_{N-1,s}
 ~~=~~
 \sum_{k\in\bbZ}
 \frac{q^{k(s+1)}(1+q^{k+N})}{D_N(q^k)} ~~=~~
 T_{N,s}+q^N \cdot T_{N,s+1}. 
\end{equation}
Similarly, since $D_N(t)=(1+t) \cdot D_{N-1}(qt)$, we have
\begin{align}
 T_{N,s}+T_{N,s+1}
 ~~&=~~
 \sum_{k\in\bbZ}
 \frac{q^{k(s+1)}(1+q^k)}{D_N(q^k)}
 \notag\\
 ~~&=~~
 \sum_{k\in\bbZ}
 \frac{q^{k(s+1)}}{D_{N-1}(q^{k+1})}
 \notag\\
 ~~&=~~
 q^{-(s+1)} \cdot
 \sum_{\ell\in\bbZ}
 \frac{q^{\ell(s+1)}}{D_{N-1}(q^\ell)}
 \notag\\
 ~~&=~~
 q^{-(s+1)} \cdot T_{N-1,s},
 \label{eq:sum-recursion-two}
\end{align}
where the third equality follows by substituting $\ell=k+1$.

Combining equations~\eqref{eq:sum-recursion-one}
and~\eqref{eq:sum-recursion-two}, we find that $T_{N,s}+q^N \cdot T_{N,s+1} =
 q^{s+1} \cdot (T_{N,s}+T_{N,s+1})$, and hence $ (q^{s+1}-q^N) \cdot T_{N,s+1}
 =
 (1-q^{s+1}) \cdot T_{N,s}$. Since $s\le N-2$ and $0<q<1$, the coefficient on the left-hand side of this equality
is strictly positive. Dividing by this coefficient and factoring
$q^{s+1}-q^N=q^{s+1}(1-q^{N-s-1})$, we obtain
\[
 T_{N,s+1}
 ~~=~~
 \frac{1-q^{s+1}}{q^{s+1}-q^N} \cdot T_{N,s}
 ~~=~~
 q^{-(s+1)} \cdot 
 \frac{1-q^{s+1}}{1-q^{N-s-1}} \cdot T_{N,s}.
\]

\end{document}